\documentstyle[epsfig,amssymb,onecolumn]{mn2e}

\newif\ifAMStwofonts

\title[Warps and the intergalactic medium]{On the origin of warps 
and the role of the intergalactic medium}
\author[F. J. S\'anchez-Salcedo]{F. J. S\'anchez-Salcedo\thanks{E-mail:
jsanchez@astroscu.unam.mx}\\
Instituto de Astronom\'{\i}a, Universidad Nacional
Aut\'onoma de M\'exico, Ciudad Universitaria,\\
Apt.~Postal 70 264, C.P. 04510, Mexico City, Mexico}
\begin{document}

\date{Accepted xxxx Month xx. Received xxxx Month xx; in original form 
2004 October 5}

\pagerange{\pageref{firstpage}--\pageref{lastpage}} \pubyear{2002}
\maketitle

\label{firstpage}

\begin{abstract}
There is still no consensus as to what causes galactic discs to become
warped. Successful models should account for the frequent occurrence of 
warps in quite isolated galaxies, their amplitude as well as the
observed azimuthal and vertical distribution of the H\,{\sc i} layer.
Intergalactic accretion flows and intergalactic magnetic fields
may bend the outer parts of spiral galaxies. In this paper
we consider the viability of these non-gravitational torques to take
the gas off the plane. We show that magnetically generated warps are
clearly flawed because they would wrap up into a spiral in less than
two or three galactic rotations. The inclusion of any magnetic diffusivity
to dilute the wrapping effect, causes the amplitude of the warp to damp.
We also consider observational consequences of the accretion of
an intergalactic plane-parallel flow at infinity.
We have computed the amplitude and warp asymmetry in the accretion
model, for a disc embedded in a flattened dark matter halo, 
including self-consistently
the contribution of the modes with azimuthal wavenumbers $m=0$ and $m=1$.
Since the $m=0$ component, giving the U-shaped profile, is not 
negligible compared to the $m=1$ component, this model predicts 
quite asymmetric warps, maximum gas displacements
on the two sides in the ratio $3:2$ for the preferred Galactic parameters, 
and the presence of a fraction $\sim 3.5\%$ of
U-shaped warps, at least. 
The azimuthal dependence of the moment transfer by the ram pressure
would produce a strong
asymmetry in the thickness of the H\,{\sc i} layer and asymmetric
density distributions in $z$, 
in conflict with observational data for the warp in our Galaxy and in external
galaxies.  The amount of accretion that is required to explain the Galactic
warp would give gas scaleheights in the far outer disc that are too small. 
We conclude that accretion of a flow with no net angular momentum,
cannot be the main and only cause of warps.
\end{abstract}


\section{Introduction}
The generation and persistence of galactic warps have been a
puzzle for galactic dynamicists over the last decades
(see Toomre 1983; Binney 1991, 1992 for reviews)
and it is still a problem of huge activity. Intergalactic winds,
tidal interactions with satellites, magnetic fields, misaligned
dark matter haloes, cosmic infall, vertical resonances and
bending instabilities have all
been proposed as causative agents (e.g., Binney 1992; Griv et al.~2002;
Revaz \& Pfenniger 2004).
Understanding the underlying physics 
responsible for the generation of galactic warps may shed light on
the late stages of accretion in galaxies or the interaction between 
the periphery of galaxies and the teneous intergalactic medium. 
Motivated by some recent observational correlations that favour
scenarios in which the intergalactic medium is a necessary ingredient
to understand the warp mechanism (Castro-Rodr\'{\i}guez et al.~2002;
S\'anchez-Saavedra et al.~2003), we have reconsidered the role
of the intergalactic medium.

The idea that the Galactic gas layer bends by a subsonic (intergalactic) 
wind was proposed by Kahn \& Woltjer (1959). In this picture, which we
refer to as the wind model,  
the discs of the galaxies are distorted because of pressure differences
exerted by a wind flowing past the ``solid'' halo of hot gas. No accretion of
mass in the disc occurs in this model because the coronal 
gas acts as an obstacle.
Saar (1979) shown that if the coronal gas responses against the
colliding wind as a smooth halo,
which is a better representation than a rigid obstacle, the warping effect
is considerably reduced. A major problem with the wind model is that
the required massive wind, for which there is no evidence, 
should be observable (Binney 2000). 

L\'{o}pez-Corredoira et al.~(2002), hereafter referred to as LC, 
have explored in some detail the idea proposed by Mayor \& Vigroux (1981) 
that the accretion of baryonic matter, probably in the form of High
Velocity Clouds (HVCs), as the galaxy moves through
the intergalactic medium, can explain the existence of warps.
Unlike the wind model, the intergalactic gas flow is supposed
to be accreted directly onto
the disc creating a ``collisional torque''. We will refer to the latter
scenario as the accretion model.

\begin{figure}
\epsfig{file=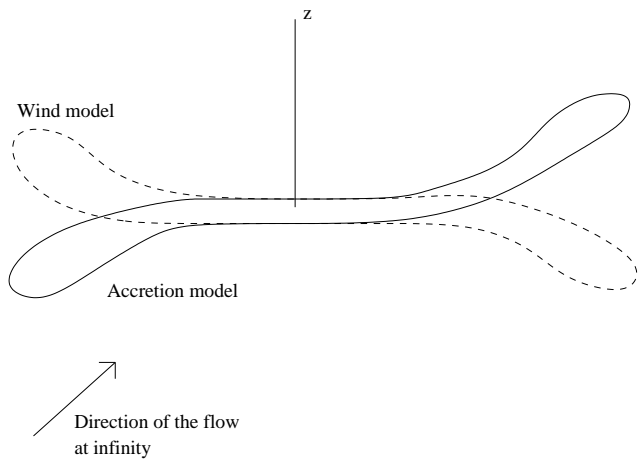,
        angle=0,
        width=3.3in
}
\caption{Schematic diagram showing the shape of the warp for the wind 
model and for the accretion model. }
\label{fig:wversusa}
\end{figure}

The wind and accretion models have in common that the impulse transfer
in an oblique incident flow, has an azimuthal dependence and, hence,
it can act as a source of excitation of warps. However, the azimuthal
dependence of the impulse transfer is not the same for both models.
For a flow oriented as in Fig.~\ref{fig:wversusa}, 
the wind model predicts that the left side of the galaxy would lie 
above the plane defined
by the inner galaxy and the right side should be placed below this
plane (see, e.g., Ikeuchi \& Tomisaka 1981).
However, in the accretion model, it holds the opposite.
Therefore, since their effects should partially counteract,
the amplitude of the warp depends on what fraction of the intergalactic
gas resides in clouds that can reach the disc without evaporating.

It is also important to remind that the accretion model is
absolutely different to the cosmic infall one
(e.g.~Ostriker \& Binney 1989). These authors pointed out that
the accretion of spherical shells of protogalactic material
in the {\it outer halo} shifts significantly the direction of
the net angular-momentum vector of a typical galactic halo
every Hubble time and that the resulting {\it gravitational torques} 
are liable to warp the disc. 

An alternative model that invokes the intergalactic medium as a potential
source to induce warps was provided by Battaner et al.~(1990). These authors 
suggested that the tension of galactic magnetic fields that arises
in the periphery of galactic gas discs when the field lines exit the
disc and connect with the uniform intergalactic field,
produces the required torque. This magnetic
model predicts that the gas layer should bend more strongly than the
stellar disc,
which motivated Cox et al.~(1996) to observe the galaxy UGC 7170
in both optical and radio (see also Casertano et al.~1987). 
They did not see any noticeable difference
between them, but a single object is not enough to rule out the magnetic
hypothesis in general. On the other hand, Florido et al.~(1991) 
found a clear wavelength dependence in the warp's curve for three galaxies.
This fact could suggest that warps are induced by a force acting on the gas
but not on stars.

A virtue attributed to these models (the magnetic scenario, the wind
and the accretion model) is that they can explain
the alignment of warps of different galaxies 
in a natural way (e.g., Battaner et al.~1991).
Other advantage of these models is 
that they work with no need for the existence
of a massive dark halo of collisionless particles
(see LC; Castro-Rodr\'{\i}guez et al.~2002). 

The first question that arises
is whether warps trace fairly the cosmic infall or, on the
contrary, intergalactic accretion flows and
magnetic forces are working simultaneously, either
coherently or destructively, as bending mechanisms. 
In the latter case it would be interesting to constrain the contribution
of each agent.
It seems unlikely that all the different warping mechanisms
may be operating simultaneously with
comparable strength. 
It is important to ascertain the role of the different torques in
modelling the outer parts of disc galaxies,
being aware of the difficulties inherent to each scenario.
In this paper we will analyse the effects of the non-gravitational
torques. In \S 2 we provide the departure point for our study of the
steady-steate configurations of gas discs under the action of torques.
Section 3 is devoted to the
magnetic model, whereas the accretion model is defered to \S 4.
We conclude that the effect of non-gravitational torques, either
magnetic or hydrodynamical by the intergalactic medium, is not the
main cause for warps generation in field large spirals. 
Exceptionally, cluster galaxies suffering from strong ram pressure 
stripping could be bent by dynamical pressure.

\section{Torques on galactic discs. Basic considerations}
\label{sec:basic}
Kahn \& Woltjer (1959) shown that, as galactic warps cannot be primordial
because of the so-called winding problem,
an external torque to take the gas off the plane should be required. 
These torques can be due to tidal forces, magnetic fields,
by ram pressure or by cosmic infall of matter in the outer halo.
We must warn that the presence of torques is not guarantee of
the existence of an immortal steady antisymmetric warp. It is essential to study
the evolution of the disc to ensure that the initial forced integral-sign
form of the warp is not dispersed. 
In this Section we will discuss some basic points on the evolution
of a gas disc subject to torques, which are necessary
to understand the remainder of the paper.

\subsection{External torques and coordinate system}
\label{sec:externaltorques}
The variation of the angular momentum of the
whole galactic disc $\bmath{L}_{\rm disc}$ (stars plus gas) will depend only
on the {\it external} torques according to
the equation:
\begin{equation}
\frac{d\bmath{L}_{\rm disc}}{dt}=\bmath{\tau}_{\rm h}+
\bmath{\tau}_{p}+\bmath{\tau}_{\rm ng},
\end{equation}
where $\bmath{\tau}_{\rm h}$ is the integrated torque over the disc
produced by the dark matter halo,
$\bmath{\tau}_{p}$ the gravitational torque due to the presence of a 
perturber (as a companion galaxy) and 
$\bmath{\tau}_{\rm ng}$ is the torque caused by a non-gravitational force,
$\bmath{f}_{\rm ng}$, which may have a magnetic origin  
(see \S \ref{sec:magnetic}) or may describe the action of the dynamical 
pressure applied by the intergalactic medium onto the disc
(\S \ref{sec:interflow}). 
In the present work we investigate the role
of the force $\bmath{f}_{\rm ng}$ in shaping the warp of galaxies 
in tidal isolation such that $\bmath{f}_{p}=\bmath{\tau}_{\rm p}=0$ 
is a good approximation.

For a galaxy embedded in a spherical halo we have
$\bmath{\tau}_{h}=0$, hence
\begin{equation}
\frac{d\bmath{L}_{\rm disc}}{dt}= \bmath{\tau}_{\rm ng}. 
\end{equation}
So, in a stationary configuration, the disc must precess with
a no-null frequency $\omega$.
Nevertheless, for the models considered in this paper, $\omega$ is 
very small (see \S \ref{sec:caveats} and LC). 
In particular, the frequency $\omega$ in both the accretion and
magnetic models is of $\sim 1$ cycle per $32$ Gyr. This period is much larger
than the characteristic dynamical time-scale in the disc, $\lesssim 1$ Gyr,
and reflects the fact the the disc is very reluctant to precess as a whole.
Once the precession of the disc is neglected,
the inner or central disc (up to $\sim 10$--$12$ kpc in the Galaxy), 
which is almost unaffected by the external torque,
defines our inertial coordinate system, being the $z$-axis
perpendicular to the central disc. The error made by this approximation
in the calculation of torques in the warp is ${\mathcal{O}}(\omega/\Omega)$,
where $\Omega$ is the angular speed.
This approximation allows us to take the inner disc as being non-responsive
so that the gravitational force
due to the inner disc is incorporated in the model as an external background
force acting on the outer disc, and consider only the evolution of the
the outer parts of the galaxy, where
the surface density of the disc drops quickly with radius and the
disc starts to bend. The equation for the angular
momentum of the outer disc, $\bmath{L}^{(\rm out)}_{\rm disc}$, is  
\begin{equation}
\frac{d\bmath{L}^{(\rm out)}_{\rm disc}}{dt}= \bmath{\tau}_{\rm ng}^{(\rm out)}
+\bmath{\tau}_{\rm inn}^{(\rm out)}, 
\end{equation}
where $\bmath{\tau}_{\rm inn}^{(\rm out)}$ is the torque produced by the
inner disc onto the outer disc.
If, for some reason, it holds $\bmath{\tau}_{\rm ng}^{(\rm out)}
=-\bmath{\tau}_{\rm inn}^{(\rm out)}$, 
then the mean precession of the outer disc will be zero. However, if we model
the outer disc as a collection of concentric rings and since galactic discs
do not behave as rigid solids, each ring may precess at a
different frequency, even though the external torques cancel out. 

On the other hand, if the dark halo is oblate or prolate and if it is taken
to be frozen so that the effect of the disc gravity
on the halo is thus neglected, a situation 
$\bmath{\tau}_{\rm h}=-\bmath{\tau}_{\rm ng}$ is possible, 
implying that $d\bmath{L}_{\rm disc}/dt=0$.
In that case the mean precession of the outer disc vanishes, 
as external torques are in balance.

\subsection{Equation of motion and bending modes}
It is convenient to make discussions in terms of the vertical
displacement of the disc at $(R,\phi,t)$, denoted by $h(R,\phi,t)$,
where $R$ is the perpendicular distance from the $z$-axis
and $\phi$ is the azimuthal angle measured in the equatorial plane.
As it is usual (e.g. Hunter \& Toomre 1969; Sparke \& Casertano 1984),
we will assume throughout this paper that the vertical scale
is large enough compared to the horizontal epicycle that the 
epicyclic motion in $R$ and $\phi$ can be neglected.
The entire vertical dynamics can be summed up by the equation
\begin{equation}
\left(\frac{\partial}{\partial t}+\Omega\frac{\partial}{\partial \phi}
\right)^{2}h=f_{\rm self}+f_{h}+f_{\rm ng},
\label{eq:hequation}
\end{equation}
where $f_{\rm self}(R,\phi,t)$ and $f_{h}(R,\phi,t)$ 
represent gravitational vertical forces per unit mass due to the 
self-interaction of the disc
material and interaction with the dark halo, respectively. 
$f_{\rm ng}(R,\phi,t)$
is a non-gravitational force (per unit mass) to be 
specified, and $\Omega (R)$ is the angular speed about the axis $R=0$.
In practice, since the disc is assumed to
stay close to the midplane $z=0$, i.e.~$h\ll R$, $f_{\rm ng}$ does not
depend on $h$ to first order.

To first order in $h/R$, the vertical force $f_{\rm self}(R,\phi,t)$ due
to the disc itself is given by
\begin{equation}
f_{\rm self}=G\int_{0}^{\infty} \Sigma(R') R' dR'\int_{0}^{2\pi}
d\phi' \frac{h(R',\phi',t)-h(R,\phi,t)}{\Delta (R,R',\phi,\phi')},
\label{eq:selfcomplete}
\end{equation}
where $\Delta (R,R',\phi,\phi')=(R^2 +R'^{2}-2RR'\cos(\phi-\phi')+
z_{0}^2)^{3/2}$ and $z_{0}$ is a softening radius that arises
because the disc has finite vertical thickness. 
If the dark halo is axially symmetric about $R=0$ and mirror
symmetric about the plane $z=0$, $f_{h}$ can be approximated by
\begin{equation}
f_{h}(R,\phi,t)=-\nu_{h}^{2} h(R,\phi,t),
\end{equation}
where $\nu_{h}^{2}(R)\equiv \partial^{2} \Phi_{h}/\partial z^{2}$
evaluated at the equatorial plane, with $\Phi_{h}(R,z)$ the gravitational
potential created by the dark halo.
After specifying $f_{\rm ng}$ and the initial conditions of the
disc, the subsequent motion of the disc is completely described
by Eq.~(\ref{eq:hequation}). 

First, we will focus on modes with azimuthal wavenumber $m=1$ of the form 
$h(R,\phi,t)=h_{1}(R)\sin(\phi-\omega_{R} t+\eta)$, where 
the eigenfrequency $\omega_{R}$ may depend on $R$, and $\eta$ is
a phase which could depend also on $R$ but it is chosen constant 
in order to have an integral sign warp at $t=0$.
Each mode consists of a collection of concentric tilt rings
that precess with frequency $\omega_{R}$.
If the frequencies of precession $\omega_{R}$ vary between
successive rings, the 
line of nodes, where material at each radius encounters the reference
galactic plane, will wrap in a spiral form in a characteristic time 
$\sim \pi/[\omega_{R}(R_{1})-\omega_{R}(R_{2})]$, where $R_{1}$ and
$R_{2}$ are the minimum and maximum radii of the warp. 
In order to avoid the winding problem,
we are mainly interested in modes in which the galaxy
can precess as a reasonably rigid unit or, conversely,
that $\omega_{R}$ is approximately constant in $R$.

If $f_{\rm ng}=0$, the gravitational torques exerted in each
ring drive them to precess with $\omega_{R}\neq 0$ about $R=0$.
It is straightforward to estimate $\omega_{R}$ in the simplest approximation
that the gas in the outer disc moves as test particles in a fixed
axisymmetic galactic potential, then $f_{\rm self}=-\nu_{d}^{2}h$.
Equation (\ref{eq:hequation}) is simplified to
\begin{equation}
\left(\frac{\partial}{\partial t}+\Omega\frac{\partial}{\partial \phi}
\right)^{2}h=-\nu_{t}^{2}h,
\label{eq:freetorque}
\end{equation}
where $\nu_{t}^{2}(R)\equiv\nu_{d}^{2}(R)+\nu_{h}^{2}(R)$. 
After direct substitution of $h$ 
into Eq.~(\ref{eq:freetorque}), we find that the rings precess with
eigenfrequencies $\omega_{R}=\Omega\pm\nu_{t}$.
For an oblate potential $\nu_{t}>\Omega$,  
so the rings regress with angular frequency $\nu_{t}-\Omega$, or
advance with frequency $\nu_{t}+\Omega$.  
Kahn \& Woltjer (1959) soon noticed that
for a galaxy like the Milky Way, the warp will
wind up in a characteristic time $<2$ Gyr, which is very short compared
to the age of the disc (see also Binney 1992). 

Sparke \& Casertano (1988) shown that if one uses Eq.~(\ref{eq:selfcomplete})
instead of the approximation $f_{\rm self}=-\nu_{d}^{2}h$, 
a self-gravitating disc
embedded in a flattened dark halo can support a discrete tilted mode
since it is able to reconcile the divergent precession rates of inner and
outer rings. However, Nelson \& Tremaine (1995) found that when
the response of the dark halo to the precession of the disc is included,
the modified tilted mode is damped within a few dynamical times
(see also Binney et al.~1998).

In the next Subsection, we will study the bending waves of a disc 
which is subject to an extra force $f_{\rm ng}$. 
The existence of modes in which the disc can precess as a whole depends 
on the form of $f_{\rm ng}$. 
As $f_{\rm ng}$ does not necessarily
depend linearly on $h$, we refer to $f_{\rm ng}$ as the ``forcing term''
of the complete inhomogeneous equation (\ref{eq:hequation}). 
When $f_{\rm ng}\neq 0$, the solution will be a superposition of
the above-mentioned modes arising
from the homogenous equation ($f_{\rm ng}=0$) plus the particular solution.
Since those waves associated with the homogeneous equation 
are unable to survive as long-lived configurations,
we will be only interested in the particular solution.

\subsection{Stationary configurations}
\label{sec:basic22}

For reasons that will become clear later, we will consider two cases: 
case A and case B. In case A, we assume that the force 
$f_{\rm ng}$ does not depend on time, i.e.~$f_{\rm ng}=f_{\rm ng}(R,\phi)$.
Since we wish to study only distortions with
wavenumber $m=1$, it is natural
to restrict the present analysis to a vertical force with $m=1$.
Without the loss of generality, this is accomplished by a force
of the form $f_{\rm ng}(R,\phi)=f_{1}(R)\sin\phi$, where
$f_{1}(R)$ represents an arbitrary function of $R$.
Case B assumes that the force is described by 
$f_{\rm ng}(R,\phi,t)=f_{1}(R)\sin (\phi-\Omega t)$.
Case A and B are relevant for the wind and magnetic 
models, respectively (see \S\S \ref{sec:magnetic} and \ref{sec:interflow}).

{\it Case A.-} 
Adding the vertical force, the dynamical equation for $h$ is
\begin{equation}
\left(\frac{\partial}{\partial t}+\Omega\frac{\partial}{\partial \phi}
\right)^{2}h=-\nu_{t}^{2}h+f_{1}(R)\sin\phi.
\label{eq:consttorque}
\end{equation}
Here, for the sake of clarity, we have assumed that $f_{\rm self}=
-\nu_{d}^{2}h$
(see \S \ref{sec:interflow} 
for a more realistic approximation for $f_{\rm self}$).
The solution of this equation will be the sum of the 
solution of the homogeneous equation (Eq.~\ref{eq:freetorque}), 
plus the particular solution $h_{p}$ given by
\begin{equation}
h_{p}(R,\phi ,t)=\frac{f_{1}(R)}{\nu_{t}^{2}-\Omega^{2}}\sin \phi, \,\,\,\,\,
(\nu_{t}\neq\Omega).
\end{equation}  
Restricting ourselves to modes with azimuthal wavenumber $m=1$,
there are three azimuthal frequencies for vertical excitations:
$\nu_{t}-\Omega$, $\nu_{t}+\Omega$ and $0$ (always present).
For arbitrary initial conditions the evolution is more complex.
For illustration, in Appendix A we provide the evolution of a disc that 
initially lies in the equatorial plane with zero vertical velocity.
As already said, we will retain only the particular solution as those bending
waves associated with the homogenous equation will disappear rapidly 
as a result of the winding process aided by dissipation due to
cloud-cloud collisions \footnote{In other words,
we are not interested in the initial-value problem but in the
existence of a mode in which the galaxy can precess as a reasonably
rigid unit. This is very common in other areas
of physics as when dealing with the propagator of particles
in the Feynman diagrams, where a pole is introduced in order
to have a well-defined momentum before and after interactions.}.

The particular solution, which is independent of the initial
conditions and the only observable distortion
at later times, is a zero-frequency mode (rings do not precess at all)
with a straight line of nodes aligned with the $x$ axis.
The orbits tilt around the $y$ axis towards the $x$ axis
only close to the resonance 
$\nu_{t} \approx \Omega$, condition which is not met in galactic
discs. For a description of the particular solution in terms of orbits,
we refer the reader to Appendix A. Essentially,
a cloud under the action of a periodic vertical force executes a motion
along $z$ which is a combination of two oscillations, one with
the intrinsic frequency $\nu_{t}$ and one with frequency $\Omega$ of the
force $f_{\rm ng}$. All we say is that if the amplitude of the 
intrinsic mode is zero, the cloud moves with the same frequency 
$\Omega$ in the three directions $x$, $y$ and $z$, 
thereby describing a closed circular orbit in a
plane inclined to its galaxy's equatorial plane and rotated around the
$x$-axis. 

The stationary solution 
$h(R,\phi)=(\nu_{t}^{2}-\Omega^{2})^{-1}f_{1}(R)\sin\phi$
describes a collection of tilt rings which do not precess.
One could wrongly think that a situation with $\omega_{R}=0$
is not possible because each ring is subject
to an external torque and hence all the rings must necessarily precess.
{\it There is no any inconsistency;
the total (disc$+$halo$+$applied by $f_{\rm ng}$) torque 
acting on each ring is zero in this particular
configuration} (see Appendix A for details).
In fact, integrated over a ring, the force $f_{\rm ng}$ produces
a torque directed around the $x$-axis, but this torque is balanced
by the torque created by the gravitational restoring forces
$f_{\rm self}+f_{h}$.

In conclusion, for a forcing term of the type 
$f_{\rm ng}(R,\phi)=f_{1}(R) \sin\phi$ 
the only long-lived warped configuration has a zero precession rate.

{\it Case B.-} We consider now a case in which the non-gravitational
force can be expressed as $f(R,\phi,t)=f_{1}(R)\sin(\phi-\Omega t)$.
With the approximation $f_{\rm self}=-\nu_{d}^{2}h$,
the corresponding equation for $h$ is
\begin{equation}
\left(\frac{\partial}{\partial t}+\Omega\frac{\partial}{\partial \phi}
\right)^{2}h=-\nu_{t}^{2}h+f_{1}(R)\sin\left(\phi-\Omega t\right).
\label{eq:prectorque}
\end{equation}
The non-gravitational force that feels a certain cloud 
is independent of time since 
both the cloud and the force rotate around the $z$-axis with the same
angular frequency $\Omega$.

Following the same reasoning than in case A,
only the particular solution given by
\begin{equation}
h_{p}(R,\phi, t)=\frac{f_{1}(R)}{\nu_{t}^{2}}\sin\left(\phi-\Omega t\right),
\label{eq:caseB}
\end{equation}
is considered. The corresponding vertical velocity
of the fluid for this solution is 
\begin{equation}
v_{z}\equiv \left(\frac{\partial}{\partial t}+\Omega \frac{\partial}
{\partial \phi}\right)h=0.
\label{eq:defvz}
\end{equation}
According to Eq.~(\ref{eq:caseB}), the vectors normal to the rings 
precess about the $z$ axis
with frequency $\Omega(R)$ and, since the frequencies
of these modes differ for successive rings in a disc with differential
rotation, their relative phases will evolve in time and the bending
packet will disperse in a characteristic time
$\sim \pi/[\Omega(R_{1})-\Omega(R_{2})]$, where $R_{1}$ and 
$R_{2}$ are the minimum and maximum radii of the warp. For a typical
galactic disc with
a flat rotation curve, this time is even shorter than that for
those modes arising from the homogeneous solution.
Consequently, this form of forcing cannot handle a long-lived warp. 

Sparke \& Casertano (1988) shown that an isolated (i.e.~$f_{\rm ng}=0$) 
self-gravitating disc
embedded in a flattened halo can support a discrete tilted mode
since it is able to reconcile the divergent precession rates of inner and
outer rings.
If $h_{\rm SC}(R,\phi,t)$ is the Sparke \& Casertano normal mode
of oscillation with $f_{\rm ng}=0$, then 
the solution of the dynamical equation for $m=1$ waves, after adding the
forcing term, has the form
\begin{equation}
h(R,\phi,t)=h_{\rm SC}(R,\phi,t)+h'_{p}(R)\sin(\phi-\Omega t),
\end{equation}
where $h'_{p}$ satisfies the integral equation
\begin{eqnarray}
&&f_{1}(R)-\nu_{h}^{2}h'_{p}(R)=Gh'_{p}(R)\int_{0}^{\infty}
\Sigma (R') H(R,R') R'dR'\nonumber\\
&&-G\int_{0}^{\infty}\Sigma(R') J(R,R')h'_{p}(R')R'dR',
\label{eqn:scf}
\end{eqnarray}
with
\begin{equation}
H(R,R')=\int_{0}^{2\pi}\frac{d\phi}{(R^{2}+R'^{2}-2RR'\cos\phi+z_{0}^{2})^{3/2}},
\label{eq:Hmayuscula}
\end{equation}
\begin{equation}
J(R,R')=\int_{0}^{2\pi}\frac{\cos\phi\, d\phi}{(R^{2}+R'^{2}-
2RR'\cos\phi+z_{0}^{2})^{3/2}}.
\label{eq:Jmayuscula}
\end{equation}
The discrete normal mode $h_{\rm SC}$ will be damped within
one dynamical time of the disc by dynamical friction
with halo particles (Nelson \& Tremaine 1995), 
whereas the bending waves due to forcing, being of the form
$\sin(\phi-\Omega t)$ with $\Omega(R)\propto 1/R$, will also
dissapear quickly by the wrapping process. Therefore, even
including self-gravity, a torque of type B cannot contribute to
alleviate the problem of the persistence of warps.

\section{Galactic warps in the magnetic scenario}
\label{sec:magnetic}
Battaner et al.~(1990) pointed out that extragalactic field lines
entering the outer parts of the gas discs could lead to a
tension able to produce a torque. 
The attractive points of this model have been stressed by different
authors (Battaner et al.~1991; Kuijken 1991;
LC; Castro-Rodr\'{\i}guez et al.~2002).
If this mechanism is at work, warps could trace out the
direction of ordered extragalactic magnetic fields at Mpc scales
(e.g., Vall\'ee 1991). LC argue that a field 
strength of $\sim 1.4$ $\mu$G could explain
the warp's curve of the Milky Way very well and, therefore, this mechanism
should not be taken lightly. Since observational determinations
of the strength of intergalactic magnetic fields are very difficult
and scarce,
it would be desirable to make more theoretical predictions and compare
with observations to quantify how much magnetic fields can contribute
to support warps. In Section \ref{sec:diffrot}, we will show that 
magnetically driven $m=1$ distortions will wrap in a spiral in
a short time-scale due to the stretching by differential rotation. 
Turbulent or ambipolar diffusion can be invoked to dilute the
undesirable wrapping effect. In that case, however, the amplitude
of the warp will decay in a shorter time-scale (\S \ref{sec:tdiffusion}).
Before doing so, we will briefly review the hypothesis behind the
magnetic scenario for warps formation.

For our purposes we only need to consider the vertical component of the 
equation of motion of a gas disc embedded in a general large-scale 
magnetic field:
\begin{eqnarray}
&&\rho\frac{d v_{z}}{dt}=-\frac{\partial P}{\partial z}
-\rho \frac{\partial \Phi}{\partial z} \nonumber\\
&& +\frac{1}{8\pi}\frac{\partial B_{z}^{2}}{\partial z}+
\frac{B_{R}}{4\pi}\frac{\partial B_{z}}{\partial R}+
\frac{B_{\phi}}{4\pi R}\frac{\partial B_{z}}{\partial \phi}.
\label{eqn:verticalMAG}
\end{eqnarray}
Here $d/dt$ is the covariant derivative and $P=P_{t}+P_{m}$, with
$P_{t}$ the gas pressure consisting of the turbulent and thermal components
and $P_{m}$ is the magnetic pressure. Consider now the third term of the 
RHS of Eq.~(\ref{eqn:verticalMAG}). The total vertical force at $(R,\phi)$,
due to this term, integrated along the vertical direction of the layer 
between $z=-z_{0}$ and $z_{0}$, with $z_{0}$ much larger
than the typical scaleheight of the disc, should be very small
or zero as $\int_{-z_{0}}^{z_{0}}\partial B_{z}^{2}/\partial z\,dz\approx 0$
for a disc embedded in a homogeneously magnetized ambient medium. 
Therefore, this term could contribute to flare the disc but
is unable to generate either a net force or a torque in the disc.
A similar argument holds for the pressure gradient term.
Hence, integrated over all $z$, the net vertical acceleration of the disc is
\begin{equation}
\left(\frac{\partial}{\partial t}+\Omega\frac{\partial}{\partial \phi}
\right)^{2}h=-\nu_{t}^{2}h+
\frac{B_{R}}{4\pi \hat{\rho}}\frac{\partial B_{z}}{\partial R}+
\frac{B_{\phi}}{4\pi \hat{\rho}R}\frac{\partial B_{z}}{\partial \phi},
\label{eq:magtorque}
\end{equation}
where $\hat{\rho}\equiv \Sigma/H$, with $\Sigma(R)$ the 
disc's surface density of gas and $H(R)$ the thickness of the disc.
For simplicity, we have adopted the simplest assumption 
that the potential created by the disc and the dark halo is fixed.

\begin{figure}
\epsfig{file=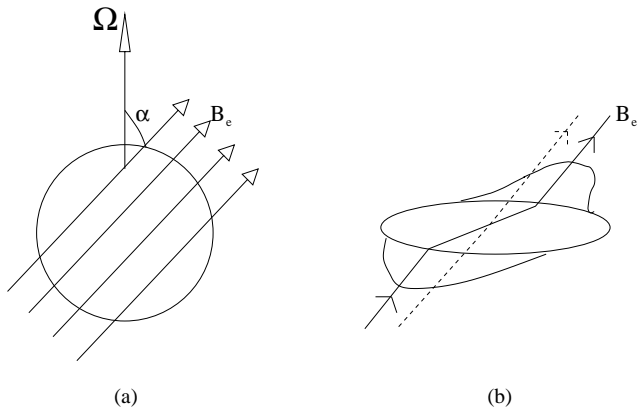,
        angle=0,
        width=3.3in
}
\caption{
Vertical collapse of a protogalaxy as an illustration of the
formation of the warp in the magnetic scenario. 
(a) The protogalaxy before collapse, 
showing a uniform intergalactic field threading it.
The magnetic field makes an angle $\alpha$ with the rotation axis
$\Omega$.
(b) The magnetic field lines (solid line), together with the warp,
after collapse to the galactic disc.
A more detailed description of intermediate stages can be found
in Howard \& Kulsrud (1997).
During the time when the galaxy contracts we ignore any rotation. 
The turbulent diffusion acts to straighten out the magnetic
field lines (dashed line), leading to the reduction of the
magnetic tension (see \S \ref{sec:tdiffusion}).}
\label{fig:howard}
\end{figure}

In the magnetic model for the formation of warps, the gas disc
is bent in an integral-sign warp by the term 
$B_{R}\partial B_{z}/\partial R$ (Battaner et al.~1990).
The reason is as follows. The extragalactic magnetic field in
the open space, denoted by $\bmath{B}_{e}$, 
is assumed to be uniform and plane-parallel, forming
an angle $\alpha$ with the plane of the galactic central disc.
Without loss of generality, suppose for convenience that $B_{e,z}>0$.
Within the thin, inner disc (say $R<R_{1}$), the field is taken as approximately
azimuthal, with $B_{z}\approx 0$, as observations suggest, and,
therefore, the magnetic force terms in Eq.~(\ref{eq:magtorque}) 
equal zero. 
In the outer parts of the disc, the galactic field lines should connect
with the ambient intergalactic magnetic field. In order to match
the field configuration in the inner disc where $B_{z}\approx 0$
into the extragalactic magnetic field $B_{e,z}$, which is 
nonvanishing whenever $\alpha$ is not close to $0$ or to $\pi/2$,
the magnitude of the $z$-component of the magnetic field should
increase radially outwards until it connects with the extragalactic
magnetic field. Therefore  
$\partial B_{z}/\partial R$ is positive within $R_{1}<R<R_{2}$, where
$R_{2}$ is the radius of the end of the disc.
As $B_{z}$ is a consequence of the penetration of the
extragalactic field, which is supposed to depend only on $R$, $B_{z}$
is taken independent of $\phi$. For the same reason 
$B_{R}(R,\phi)=B_{R,0}(R)+B_{R,1}(R)\sin [\phi+\zeta]$, with
$\zeta$ an arbitrary constant phase. Since $\partial B_{z}/\partial R >0$,
the term $B_{R}\partial B_{z}/\partial R$ produces a torque in the
outer parts of the discs, whereas the term $B_{\phi}\partial B_{z}/\partial
\phi$ is supposed to be negligible. Since the $m=0$ mode is not
interesting as far as integral-sign shape warps concern, we will adopt
the simplest assumption $B_{R,0}=0$. In summary, the {\it initial}
magnetic field (at $t=0$) in the tilted disc takes the form:
\begin{equation}
\bmath{B}(R,\phi, t=0)=B_{R,1}(R)\sin(\phi+\zeta)\bmath{\hat{e}}_{R}+
B_{\phi}(R,\phi)
\bmath{\hat{e}}_{\phi}+B_{z}(R)\bmath{\hat{e}}_{z},
\label{eq:initialfield}
\end{equation}
with $\partial B_{z}/\partial R>0$, and $\bmath{B}\rightarrow \bmath{B}_{e}$
when $R\rightarrow \infty$. This initial magnetic configuration
naturally arises in the vertical collapse of a 
spherical protogalactic cloud with a uniform magnetic field threading it 
(see Fig.~\ref{fig:howard}), if one ignores any rotation. 
A more detailed description 
of this process can be found in Howard \& Kulsrud (1997)
and there is no need to be repeated here.
The resulting magnetic field together with the warp are sketched
in Fig.~\ref{fig:howard}b. Therefore, in order to have the intergalactic
field penetrating into the H\,{\sc i} disc, we do not have to
invoke any magnetic diffusion to transport magnetic field from
outside space into the disc.

As we have anticipated in \S \ref{sec:basic}, the presence of 
torques is not a sufficient condition for the warp to rotate as a rigid 
unit.  In order to assess the viability of this model, we need to
study the temporal evolution including the gravitational and magnetic
torques. In the original treatment of Battaner and collaborators (also
in LC), they assume
that the magnetic configuration is unaffected by differential rotation
of the disc and ignore any magnetic diffusivity as well as 
the vertical motions associated with the warp, so that 
$\partial \bmath{B}/\partial t=0$.
In the following, we will attempt to ascertain the consequences of the
inclusion of the motions of the gas. These motions alter the magnetic
field lines, and therefore, the form given in Eq.~(\ref{eq:initialfield}) 
holds only at $t=0$.

The general problem of the evolution
of a magnetized rotating disc is extremely complex. 
Nevertheless, as illustrated in
\S \ref{sec:basic}, when analyzing the success or failure of an external torque
to conduct warps, it is assumed that the solution of the homogeneous equation
gets wrapped so rapidly that in practice only the particular
solution could be observed.
The case of a magnetic torque differs from the examples in \S \ref{sec:basic}
in the fact that the magnetic field and the velocity field are not
independent variables but linked 
through the induction equation. In other words, the force depends also
on the velocity field of the fluid. In Appendix B we show that, in the
magnetic case, vertical motions can distort the magnetic configuration,
spoiling out the initial sinusoidal-form ($m=1$) of the vertical magnetic
force. The aim of this Section is not to study the evolution of a 
disc starting with a warp of arbitrary amplitude but to find a
particular solution. In fact,  
if we start with a force with azimuthal wavenumber $m=1$, we seek  
the particular solution that does not generate harmonics of higher
wavenumber. In Appendix B it is found that the particular solution
corresponds to a case where the vertical velocities are very small
or zero. Thus, we will adopt $v_{z}=0$ in the next
Subsections. The characteristic evolutionary time-scales for the magnetic
field when $v_{z}\neq 0$ are also given in Appendix B.  

\subsection{Differential rotation}
\label{sec:diffrot}
We first focus on the effect of pure differential rotation and
ignore for a moment other effects such as turbulent diffusion. 
As it will be discussed in \S \ref{sec:tdiffusion},
if turbulent diffusivity is unquenched, the role of turbulent motions may 
also have a significant influence on the evolution of the warp's amplitude.
Suppose that a disc was initially forced to warp by a magnetic torque
so that it presents a straight line of nodes. We wish to know
whether the line of nodes keeps straight or, on the contrary,
gets a spiral form as time
goes by. For this purpose we need to include the evolution of the
magnetic field. 
For frozen-in magnetic field lines, the magnetic field evolves
according to
\begin{equation}
\frac{\partial \bmath{B}}{\partial t}=
\bmath{\nabla}\times(\bmath{v}\times\bmath{B}),
\label{eq:induction}
\end{equation}
where $\bmath{v}$ is the ordered velocity of the gas, $\bmath{v}=v_{\phi}
\bmath{\hat{e}}_{\phi}+v_{z}\bmath{\hat{e}}_{z}$, and
any contribution to the large-scale magnetic field
by turbulent motions has been ignored as a first approximation.
According to the ongoing discussion, we seek the particular solution
with $v_{z}=0$. The equation of magnetic induction  with $\bmath{v}=
v_{\phi}\bmath{\hat{e}}_{\phi}=R\Omega\bmath{\hat{e}}_{\phi}$ in components is
\begin{equation}
\left(\frac{\partial}{\partial t}+\Omega\frac{\partial}{\partial \phi}\right)
B_{R}=0,
\label{eq:inductionR}
\end{equation}
\begin{equation}
\left(\frac{\partial}{\partial t}+\Omega\frac{\partial}{\partial \phi}\right)
B_{\phi}=R\frac{d\Omega}{dR}B_{R},
\end{equation}
\begin{equation}
\left(\frac{\partial}{\partial t}+\Omega\frac{\partial}{\partial \phi}\right)
B_{z}=0.
\label{eq:inductionZ}
\end{equation}
To derive Eqs.~(\ref{eq:inductionR})--(\ref{eq:inductionZ}), it was
also assumed for simplicity that the galaxies follow cylindrical rotation, 
i.e.~$\partial v_{\phi}/\partial z\approx 0$.
Following a fluid element that moves with the galactic rotation, {\it only
the azimuthal component} $B_{\phi}$ will increase due to the stretching
of the radial magnetic field by the differential rotation until it
reaches saturation by the combined effects of ambipolar and turbulent
diffusion. After integration of 
Eqs.~(\ref{eq:inductionR})
and (\ref{eq:inductionZ}) with the initial conditions 
$B_{R}=B_{R,1}(R)\sin(\phi+\zeta)$
and $B_{z}(R)$ a function only of $R$, and substituting into 
Eq.~(\ref{eq:magtorque})
with $\partial B_{z}/\partial\phi=0$, the equation of motion reads: 
\begin{equation}
\left(\frac{\partial}{\partial t}+\Omega\frac{\partial}{\partial \phi}
\right)^{2}h=-\nu_{t}^{2}h+\frac{B_{R,1}}{4 \pi \hat{\rho}}
\frac{\partial B_{z}}{\partial R}
\sin\left(\phi-\Omega t+\zeta\right).
\label{eq:magneticase}
\end{equation}
By comparing Eq.~(\ref{eq:magneticase}) with Eq.~(\ref{eq:prectorque}),
one immediately realizes that the magnetic torque corresponds to {\it case B}.
The particular solution obeys 
\begin{equation}
h_{p}(R,\phi,t)=\frac{B_{R,1}}{4\pi\nu_{t}^{2}\hat{\rho}}\frac{\partial B_{z}}
{\partial R} \sin(\phi-\Omega t+\zeta).
\end{equation}
The vertical velocity $v_{z}$ is zero in
the particular solution (see case B of \S \ref{sec:basic22}); 
this confirms our previous ad hoc assumption (see also Appendix B).
As discussed in \S \ref{sec:basic} and it becomes apparent from
the equation above, integral-sign warps cannot be maintained
because of the wrapping action of the disc's rotation. 
The reason is simple;
in a magnetically-driven warp, the line of nodes will be described 
by the locus of those elements
for which the vertical magnetic force vanishes, i.e.~$B_{R}=0$. 
Since moving with a fluid parcel, $B_{R}$ is
constant in time, the line of nodes will wrap halfway round the
galaxy in a time
\begin{equation}
\tau_{\rm nodes}=\frac{\pi}{\Omega(R_{1})-\Omega(R_{2})},
\label{eq:twarp}
\end{equation}
where $R_{1}$ and $R_{2}$ represent the minimum and maximum radii
of the warp. Evaluating Eq.~(\ref{eq:twarp}) for the Milky Way
with $\Omega(R)=v_{0}/R$, $v_{0}=220$ km s$^{-1}$,
we infer $\tau_{\rm nodes}\sim 0.32$ Gyr,
where we adopted $R_{1}\approx 9.5$ kpc and 
$R_{2}\approx 16.5$ kpc from Burton (1988). Thus, the winding
problem in this case is even more severe than in the non-magnetic case
for which the wraping time of the warp is of the
order of $\sim 2$ Gyr for a quasi-spherical halo (Kahn \& Woltjer 1959; 
Binney 1992).
It can be readily shown that the line of maximum elevation will also 
tend to wind up as the line of nodes does, 
changing the initial assumed warped appeareance
beyond recognition. The wrapping of the $R$-component of the magnetic
field, which is responsible for the winding of the warp, is a consequence
of a more general result, proved in Appendix B, that no stationary
configuration exists for a tilted disc in differential rotation
under strict frozen-in conditions.

The increasing tangled magnetic field will act as a
pressure in the vertical direction, resulting in an increase in 
the thickness of the disc, and as a tension in the radial direction
(e.g., S\'anchez-Salcedo \& Reyes-Ruiz 2004),
which could lead to a radial transport of angular momentum by
the amplified stress term $B_{\phi}B_{R}$ (e.g., Kundt 1990).
Eventually, because of the strong winding of the magnetic lines, magnetic
diffusion becomes important (Parker 1973). Hydromagnetic
waves in the disc will become nonlinear and their energy will serve to heat
the disc.  This line of arguments strongly suggests that the
contribution of a torque by the extragalactic magnetic field to a systematic
deviation from the mean galactic plane is null at the present age
of galaxies in the local Universe. 

Not all the galaxies rotate differentially. The observed H\,{\sc i}
disc of some dwarf galaxies rotate as a solid rigid $\Omega(R)=$const
as occurs in IC 2574 (Blais-Ouellette et al.~2001).
For these galaxies, magnetically-generated warps could survive. 
An obvious result of our analysis is that, even if there was a uniform
large-scale intergalactic magnetic field conductive to bend the gas discs, 
a correlation between the orientation of warps as that found
in Battaner et al.~(1991) is not expected because  
the line of nodes and the line of maximum height of the galactic warp
will rotate with angular velocity $\Omega$, different from galaxy to galaxy.
Therefore, models
aimed to study the strength and direction of intergalactic magnetic
fields that are based on possible correlations between the warps of 
different galaxies, as those given in e.g.~Vall\'ee (1991), 
are not sustained in any physical basis.

So far, only rotation has been considered. In the next Subsection we
will discuss if other effects could alleviate the magnetic winding problem.

\subsection{Turbulent diffusion and the $\alpha$-effect}
\label{sec:tdiffusion}
Interstellar gas is probably turbulent. On the one hand,
turbulent diffusion of magnetic fields can act as a mechanism
for enhancing transport rates above their molecular values.
On the other hand, turbulence together
with large-scale flows can convert small-scale
fields into large-scale fields. The simplest form of the mean field
dynamo equation that retains the basic physics is
\begin{equation}
\frac{\partial \bmath{B}}{\partial t}=
\bmath{\nabla}\times(\bmath{v}\times\bmath{B})+\bmath{\nabla}\times
(\alpha\bmath{B}-\eta_{t}\bmath{\nabla}\times\bmath{B}),
\label{eq:dynamo}
\end{equation}
where $\alpha\approx -l^{2}\Omega\cdot \nabla \ln(\rho v') 
F(\bmath{B}_{m},\Omega)$ and $\eta_{t}\approx l v' 
G(\bmath{B}_{m},\Omega)/3$, with $l$ the correlation length
of the turbulent motions $v'$, $\bmath{B}_{m}$ is the large-scale 
mean magnetic field, and $F$ and $G$ are certain quenching 
functions (e.g., Beck et al.~1996, and references therein).
The $\alpha$-term represents the effect of cyclonic motions on the mean
field.

Since the disc is warped with vertical displacement $h$, 
it is convenient to introduce the
variable $Z$ as the vertical distance to the midplane of the gas disc,
$Z\equiv z-h$. 
For simplicity, we assume that the disc is defined by sharp boundaries
at $Z=\pm H/2$ and further introduce the aspect ratio of the disc,
$\lambda=H/2R$. Specifically, we choose $H/2\simeq 1$ kpc and $R=15$ kpc,
so that $\lambda\simeq 6.5\times 10^{-2}$. 

As the disc is embedded in a constant external field, $\bmath{B}_{e}$,
we expect the same field in the upper and lower surfaces of the disc
by symmetry: 
$\bmath{B}(Z=H/2)=\bmath{B}(Z=-H/2)$. 
Given the even symmetry of the dynamo equations, 
the components $B_{\phi}$ and $B_{R}$ must be
symmetric with respect to $Z$:
\begin{equation}
B_{\phi}(Z)=B_{\phi}(-Z),
\label{eq:evenBz}
\end{equation}
\begin{equation}
B_{R}(Z)=B_{R}(-Z).
\label{eq:evenBR}
\end{equation}

The ratio between the time-scales of magnetic diffusion across the
disc and along the radius is of order of $\lambda^{2}$. This implies
that the field distribution across the disc is established relatively
rapidly, while the radial distribution evolves on a considerably
longer time-scale. Consequently, in the absence of the $\alpha$-effect, 
$B_{z}$ obeys a pure diffusion equation 
$\partial B_{z}/\partial t\simeq \eta_{t}\nabla^{2}B_{z}$, where
the magnetic diffusivity may depend on radius, $\eta_{t}(R)$.
Diffusion of the plane-parallel ambient magnetic field into the
disc can counteract the undesirable effect of differential rotation 
if the appropriate local Reynolds number 
${\mathcal{R}}_{m}=\Omega H^{2}/4\eta_{t}\lesssim 1$ (e.g., Moffatt 1978),
implying that  
the characteristic diffusion time-scale, $\tau_{D}(R)=H^{2}(R)/4\eta_{t}(R)$, 
at any radius, is smaller than local rotation
period, i.e.~$\tau_{D}(R)\lesssim 1/\Omega (R)$. In particular,
at a radius $R_{1}$, $\tau_{D}(R_{1})\lesssim 0.1$ Gyr. 
If we turn off the $\alpha$-effect, turbulent diffusion permits slip of
the field in the $z$ direction and tends to make the field more
uniform and straight (see Fig.~\ref{fig:howard}b). 
After a few local diffusion time-scales, $B_{z}(R)$ will 
take a value equals to the 
intergalactic value. Hence, the magnetic tension responsible for
the warp, which is $\propto \partial B_{z}/\partial R$, 
drops monotonically to zero in a few local $\tau_{D}$. 
Therefore, turbulent and ambipolar diffusion can delay the winding
process of magnetically-driven warps
but at the price of decreasing the amplitude of the warp
to a negligible value in a time-scale comparable to $\tau_{\rm nodes}$,
i.e.~in less than $1$ Gyr; 
the inclusion of any magnetic diffusion
aggravates the problem of the magnetic support of galactic warps. 

To show that the $\alpha$-term is not suitable to change the dramatic 
damping effect of turbulent diffusion, consider the mean vertical value
of $B_{z}$ in the disc
\begin{equation}
F_{z}(R,\phi)\equiv \frac{1}{H}\int_{-H/2}^{H/2}B_{z}\, dZ. 
\end{equation}
From the $z$-component of Equation (\ref{eq:dynamo}) 
\begin{eqnarray}
&&\frac{\partial F_{z}}{\partial t}=
\frac{1}{H}
\int_{-\frac{H}{2}}^{\frac{H}{2}}\left(\frac{1}{R}
\frac{\partial (R\alpha B_{\phi})}
{\partial R}+\eta_{t}\nabla^{2} B_{z}-\frac{1}{R}\frac{\partial (\alpha B_{R})}
{\partial \phi}\right)dZ\nonumber\\
&&=\frac{1}{H}\left(\frac{1}{R}+\frac{\partial}{\partial R}\right)
\int_{-\frac{H}{2}}^{\frac{H}{2}}\alpha B_{\phi}dZ
-\frac{1}{HR} \frac{\partial}{\partial \phi}
\int_{-\frac{H}{2}}^{\frac{H}{2}}
\alpha B_{R}dZ+\eta_{t}\nabla^{2}F_{z}+
\frac{\eta_{t}}{H}\frac{\partial B_{z}}{\partial Z}
\bigg|_{-\frac{H}{2}}^{\frac{H}{2}}
\end{eqnarray}
Because $B_{\phi}$ and $B_{z}$ are even in $Z$ (see 
Eqs.~\ref{eq:evenBz}-\ref{eq:evenBR}) and 
$\alpha$ is odd, the integrals involving the product $\alpha$ and
$B_{z}$ or the product $\alpha$ and $B_{R}$ vanish and, hence, 
\begin{equation}
\frac{\partial F_{z}}{\partial t}=
\eta_{t}\nabla^{2}F_{z}+
\frac{\eta_{t}}{H}\frac{\partial B_{z}}{\partial Z}
\bigg|_{-\frac{H}{2}}^{\frac{H}{2}}.
\label{eq:nullalpha}
\end{equation}
Equation (\ref{eq:nullalpha}) indicates that the diffusion of the
component $B_{z}$ across the cap surfaces of the disc at $Z=\pm H/2$ 
follows unimpeded even in the presence of the $\alpha$-effect.

In conclusion, in order to avoid a fast reduction of the amplitude of
the warp to unobservable values, the magnetic 
diffusivity should be significantly quenched relative to the canonical
turbulent diffusion value, $\eta_{t}$.

\subsection{The magnetic field outside the disc}
Regular magnetic fields have been detected not only in the discs of
spiral galaxies but also far above them, in the surrounding gaseous
medium. The reason is that both the magnetoionic disc, which
is thicker than the H\,{\sc i} disc, and the gaseous halo are 
magnetically active. 
Because we were only interested in the magnetic term responsible
for the warp formation and since warps are observed in the very
cold H\,{\sc i} disc, the details of the evolution of the complete
magnetic field outside the disc are not important. 
A thorough analysis of the magnetic field structure is out of the
scope of this paper. For completeness, however, we will comment on
the magnetic field outside the disc.

Initially the magnetic field was supposed to be uniform and equal
to $\bmath{B}_{e}$, the intergalactic field. The field trapped
by the galaxy will undergo the action of chaotic motions and differential
rotation. All we can say, at this point, is that the {\it outer  
boundary condition} should be that the field $\bmath{B}_{e}$ at 
infinity is undisturbed, i.e.~$\bmath{B}\rightarrow \bmath{B}_{e}$
as $r\rightarrow \infty$.

In order to study the field structure in the magnetoionic disc, we
need to implement boundary conditions at the surface of the disc
in Eq.~(\ref{eq:dynamo}). The teneous and partially ionized gas in
the corona suggests force-free conditions for the medium outside the
disc. Observations suggest that this is not the case. Coronal gas is
in chaotic motion, and acquires a nonzero mean helicity due to
galactic rotation (e.g., Sokoloff \& Shukurov 1990). 
Since the dynamo can be active not only within the disc but also
in the gaseous halo, one should resolve, in principle, the disc and
halo fields in a self-consistent calculation. 
However, the magnetic diffusivity in the halo
is $\sim 50$ times that in the disc, which indicates that it is
meaningful to consider the magnetic evolution in the galactic disc
by neglecting the dynamo in the halo and adopting vacuum boundary
conditions (e.g., Poezd et al.~1993; Bardou et al.~2001). Numerical
simulations by Moss \& Brandenburg (1992) confirm that the vacuum
boundary conditions can be applicable in this case. 
In this context, a vacuum is understood as a medium in which
electromagnetic induction effects are significantly weaker than
the magnetic diffusion.

If the disc and the corona are rotating as a rigid body, the evolution
of the regular field in the disc would be quite simple; $\alpha\simeq 0$
and diffusion effects lead to the external field $\bmath{B}_{e}$ to
penetrate into the disc; the ultimate steady-state is a uniform
field and equal to $\bmath{B}_{e}$.
Differential rotation makes the situation more complex. The response
and development of the field lines due to rotation 
is a classical problem in MHD.
The full time-dependent problem of a rotating cylinder embedded in
a perfect insulator was solved by Parker (1966). The interesting
reader is referred to this paper for a visualization of the lines
of force and its different behaviour with the Reynolds number
(see also Moffatt 1978, Section 3.8). The evolution of the field lines 
in the standard cooling flow scenario for the formation of disc 
galaxies can be found in R\"{o}gnvaldsson (1999).

\subsection{Other caveats}
In this discussion we have assumed that the center of mass of the
galaxy is at rest in the initially plane-parallel intergalactic magnetic
field, as a prerequisite for galaxies to be
integral-sign warped in the magnetic scenario.
For galaxies moving through the intergalactic medium,
shear terms will distort the field lines in the outer parts
of the disc and, therefore, the field lines cannot be considered
plane-parallel any longer, but a magnetic barrier of increased
field strength would exert a pressure on the disc
(e.g.~Konz et al.~2002), taking the classical form of a rim. 
In the next Section we will discuss whether the hydrodynamical force 
due to the accretion of intergalactic gas, which is possibly
more important than the magnetic forces in moving galaxies,
can contribute to the formation of warps.

\subsection{Comparison with previous work}
Battaner et al.~(1990) proposed the galactic
warps may have a magnetic origin since there exists 
a magnetic field distribution liable to produce a $m=1$ torque in
a slab of slightly ionized gas. 
Interestingly, the magnetic configuration
presents lines that open to the space and connects at infinity with
a uniform magnetic field,
which was identified as the ubiquitous intergalactic magnetic field.
A detailed calculation of the magnetic field structure required to
reproduce the amplitude and shape of the warp in NGC5055
was given in Battaner \& Jim\'enez-Vicente (1998).
Their analysis was carried out by considering the Lorentz force
in the vertical equation of motion and the condition 
$\bmath{\nabla}\cdot \bmath{B}=0$, although ignoring the induction
equation for the field, or more precisely, the field was assumed to
be in a steady-state configuration, i.e.~$\partial\bmath{B}/\partial t=0$.
Under these assumptions, the predictions of this model are: 
(1) the postulated intergalactic field is of order of a few $\mu$G, 
(2) a spatial correlation between
the direction of the warps in neighbouring galaxies and (3) a shift
between the stellar and gaseous warps.

However, it is clear that the magnetic field in the disc will be
advected with the flow. We have studied this process by including
the electromagnetic
induction effects in the postulated field. The assumptions of 
Battaner et al.~(1990) and Battaner \& Jim\'enez-Vicente (1998)
are only valid for slightly turbulent systems with a rigid rotation, 
such as the magnetic configuration is not destroyed
by differential rotation or turbulent diffusion.
Even for galaxies rotating as a solid body, however,
the rotation of the galactic discs excludes any alignment
between the orientation of warps of nearby galaxies by a large-scale
intergalactic field. For discs in differential rotation, 
we show that as the magnetic field evolves due to the stretching
term, warps become
twisted and destroyed in less than two or three rotation periods.
The increasing tangled
magnetic field will act as a pressure, forming a unwarped thick H\,{\sc i}
disc at later times. 
The thickness of the H\,{\sc i} disc will depend on the 
turbulent diffusion time-scale. If it is shorter than
the time-scale of winding, the warp damps its amplitude
before the disc gets thicker. 
In our analysis we have adopted a prescribed large-scale velocity
field ignoring the magnetic back reaction of the field onto
the gas dynamics, which could lead to a radial inflow of gas.
 
If galaxies travel with a peculiar velocity through a
medium permeated by an ordered magnetic field of $\sim 1$ $\mu$G, 
the magnetic barrier of increased field strength would contribute
to the ram pressure in forming a U-warp.
Therefore, we suggest that features in the morphology 
of the outer H\,{\sc i} disc can shed light on
the strength of the regular intergalactic magnetic field.

\section{Generation of warps by an intergalactic accretion flow}
\label{sec:interflow}
\subsection{Revisiting the accretion model: motivations and caveats}
\label{sec:caveats}
It is well-known that the dynamical pressure acting on a 
galaxy travelling through the intergalactic
medium (IGM) could bend the outer parts of spiral galaxies, especially
in cluster galaxies for which the interstellar gas can be even 
stripped from the galaxy (Gunn \& Gott 1972). Several galaxies in clusters
are reported in the literature whose distorted and asymmetric
extraplanar H\,{\sc i} gas is claimed to be suggestive of ram
pressure {\it stripping} by the intracluster gas, rather than accretion.
The assumptions behind the steady-state accretion model of 
small vertical perturbations, being the disc a perfect absorber of
the incident intergalactic gas, do not apply for cluster galaxies
moving supersonically in a teneous hot medium.
The aim of this Section is to test the accretion model as an
explanation of the origin of warps in relatively isolated galaxies
or galaxies in groups. Therefore we restrict our analysis to these
systems since the approximations and results derived 
in this Section cannot be extrapolated to study the extraplanar
gas components observed in cluster galaxies.

It is illustrative to estimate the effect of the ram pressure
of a wind that collides directly onto the disc. 
In order for ram pressure to bend the gas disc a height
$h$, with $h\ll R$, above the mean plane, the following order-of-magnitude
condition has to be met:
\begin{equation}
\rho_{\infty}v_{\infty}^{2}\geq \Sigma_{\rm g}\nu_{d}^{2} h,
\label{eq:gunngott}
\end{equation} 
(e.g., Gunn \& Gott 1972; Marcolini et al.~2003)
where $\rho_{\infty}$ represents the mean gas density of the IGM, $v_{\infty}$
the velocity of the galaxy through the IGM, $\Sigma_{\rm g}$ the disc's gas
surface density, $R$ the galactocentric radius, $\nu_{t}$ the
disc vertical frequency and $v_{\rm rot}$ the circular speed. 
For an exponential disc, $\nu_{d}(R)\simeq \Omega(R)\sqrt{1+22(R_{d}/R)^3}$
(Binney 1992).
For the following reference values: $v_{\rm rot}\approx 200$
km s$^{-1}$, $v_{\infty}=200$ km s$^{-1}$ (see, e.g., Bureau \& 
Carignan 2002), $R=15$ kpc and $\Sigma_{\rm g}=1$ M$_{\odot}$ pc$^{-2}$,
we require $\rho_{\infty}\geq 10^{-4}$ cm$^{-3}$ to have a displacement
of $h\approx 500$ pc. The present-day mean density of the IGM beyond
galaxy haloes in the Local Group and in small groups of galaxies 
is thought to be $\leq 10^{-5}$ cm$^{-3}$ (e.g., Murali 2000;
Bureau \& Carignan 2002; Grebel et al.~2003). With this intergalactic
volume density only $h\leq 50$ pc could be expected for the above
reference values, which seems to be too low to explain warps in large spiral
galaxies. As mentioned at the beginning of this Section, 
since the values of $\rho_{\infty}$ and $v_{\infty}$ are much larger
in clusters of galaxies,
we cannot exclude the possibility that extraplanar H\,{\sc i}
and warps seen in some galaxies 
showing signatures of ram pressure stripping are induced by
the interaction with the intracluster medium. Good candidates are the galaxies
NGC 4548 in the Virgo cluster (Vollmer et al.~1999) and HoII in
the M81 group (Bureau \& Carignan 2002; but see also Bureau et al.~2003).
To our knowledge, numerical experiments of ram pressure stripping
do not show any evidence of the formation of integral-sign warps
(e.g., Quilis, Moore \& Bower 2000; Marcolini et al.~2003;
Roediger \& Hensler 2004). 
Other possibilities for the explanation of distorted velocity fields 
in cluster galaxies have been suggested (e.g., Vollmer et al.~1999).

In the derivation of Equation (\ref{eq:gunngott}) it was implicitly
assumed that the flow behaves as a continuous gas medium. 
Interestingly, LC show that if the flow is a gas made of discrete 
collisionless undisruptive clouds, the gravitational focusing by the galactic
potential would enhance the density of the flow by an important
factor when it impacts the disc.
In fact, they obtained that the northern warp of
the Milky Way ($h\sim 2$ kpc) can be explained by the infall of matter
through a subsonic wind of $100$ km s$^{-1}$ with 
a density of $\sim 6\times 10^{-5}$ cm$^{-3}$.

For a flow approximately parallel to the rotation 
axis of the disc (face-on wind),
the shape of the outer gas disc should resemble an axisymmetric
U-rim instead of an integral-sign warp. If the flow
is oblique, the impulse transfer has an azimuthal dependence and,
hence, the vertical distortion $h$ is a combination of the $m=0$
mode, giving a U-shape profile, and the $m=1$ mode (S-shaped profile).
The success of the ring-tilted decomposition proposed by Rogstad et al.~(1974)
in reproducing the observed morphology and kinematics of real warps
suggests that the $m=1$ component is indeed dominant with respect to the
$m=0$ one.
Hunter \& Toomre (1969) noticed that in the wind model by Kahn
\& Woltjer (1959), warps should be asymmetric, with the maximum
gas displacements on the two sides of the Galaxy in the
ratio $3:2$. The similarities between the wind
model and the accretion model suggest
that the warps in the accretion model should be also asymmetric. 
Unexpectedly, LC found that the 
S-amplitude ($m=1$ mode) is much greater than the U-amplitude
($m=0$ mode) except when the flow vector is within $5^{\circ}$ 
of the spin axis of the galaxy. They claim that if all galaxies had
Milky Way parameters around $1$ in $1000$ should have a U-shaped
warp (Beckman et al.~2003). This result is very counterintuitive indeed.

\begin{figure*}
\epsfig{file=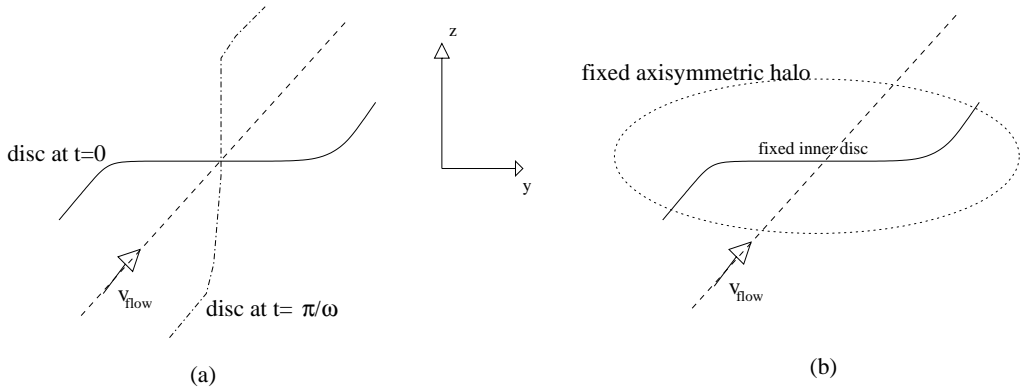,
        angle=0,
        width=5.3in
}
\caption{Graphical representation of a vertical cross section
at $x=0$ showing the warped disc.
Panel (a): Precession of the galactic disc with
frequency $\omega$ around $\bmath{v}_{\infty}$ as a single
unit (model LC). 
If the disc is embedded in the flattened potential of a massive dark halo, 
which we take to be aligned with the coordinate system (panel b), 
the direction of the incident flow is not any longer the axis of 
symmetry of the system. Both the dark halo and the central disc are
taken
as frozen and unresponsive bodies because are so massive that they are
very reluctant to precess. Contrary to panel (a), 
the outer rings cannot achieve 
large vertical heights with respect to the equatorial plane
of the dark halo ($z=0$) because the dark halo and the inner disc
try to twist the disc back into alignment with the coordinate system.
The steady-state configuration corresponds to $\omega=0$ (see text).}
\label{fig:precession}
\end{figure*}

In the stationary
models of LC\footnote{We must distinguish here between the halo of coronal 
gas and the halo made of collisionless dark matter particles. In the accretion
model, the coronal gas is ignored whereas the dark halo
may be present. The interaction of the accretion flow with the
dark halo is only through the gravitational force and, hence,
the flow can reach the disc.}, the {\it galactic disc as a whole}, and 
not only the warp, precesses as a rigid unit
with a no-null angular velocity $\omega$ around $\bmath{v}_{\infty}$,
the axis of symmetry of the system (see \S \ref{sec:externaltorques}
and Fig.~\ref{fig:precession}a). As discussed in \S \ref{sec:externaltorques}, 
since the disc is rather massive and rotates fast, $\omega$ must be  
small but different from zero.
Assuming that the dark halo is perfectly spherical, the precessing frequency 
about $\bmath{v}_{\infty}$, is given by 
$\omega^2=\bmath{\tau}_{\rm acc}^{2}/
(\bmath{L}_{\rm disc}^2\cos^{2}\theta_{0})$, 
where $\bmath{\tau}_{\rm acc}$ is the external torque acting on the disc caused
by the accretion flow of intergalactic gas and $\cos\theta_{0}=
\sin(\bmath{v}_{\infty},\bmath{L}_{\rm disc})$. 
After estimating $\bmath{\tau}_{\rm acc}$ following LC,
one gets that the precession rate around $\bmath{v}_{\infty}$
for an angle of $45^{\circ}$ is of $\sim 0.18$ km s$^{-1}$ kpc$^{-1}$ 
(see also LC).
The dynamical correction for this effect is apt to be ignored
and hence we can treat the inner disc as a fixed
external potential. Under that approximation and considering 
only the evolution of
small vertical displacements from the midplane $z=0$ defined 
by the fixed central disc, the torque applied by the flow will be
constant in direction and magnitude. This situation corresponds to
case A of \S \ref{sec:basic22} and, therefore, $\omega=0$ 
is expected in the stationary
configuration. Similarly, if we include a
massive dark halo flattened along the rotation axis and treat it as a 
time-independent background potential, the outer rings of the disc cannot
present a steady-state precession around $\bmath{v}_{\infty}$, as
occurs in a spherical halo (see the disc at $t=\pi/\omega$ in
Fig.~\ref{fig:precession}a), because the halo twists the disc back
forcing it to stay close to the halo midplane $z=0$ 
(see Fig.~\ref{fig:precession}b). 
Once the halo and central disc are treated as time-independent external
potentials, we will confirm in \S \ref{sec:massmodel} 
our expectation that $\omega=0$ in the
steady-state configuration. However, this result has a simple
explanation, as follows.

The change of the strength of the angular momentum 
of each ring, $\bmath{L}_{\rm ring}$, is 
\begin{equation}
\frac{dL_{\rm ring}^{2}}{dt}=2\bmath{\tau}_{\rm ring}\cdot\bmath{L}_{\rm ring},
\end{equation}
where $\bmath{\tau}_{\rm ring}$ is the total (gravitational plus
nongravitational) torque. In a stationary
configuration $dL^{2}_{\rm ring}/dt=0$ and hence either 
$\bmath{\tau}_{\rm ring}=0$ (and hence the rings do not precess)
or the torque is always perpendicular to $\bmath{L}_{\rm ring}$. 
As already discussed, for the time-scales of interest, 
the torque applied by the flow can be considered constant in direction
and magnitude, whereas the gravitational torques are 
always perpendicular to $\bmath{L}_{\rm ring}$.
Therefore the condition $\bmath{\tau}_{\rm ring}$ 
always perpendicular to $\bmath{L}_{\rm ring}$ cannot be met
and, in fact, the only stationary configuration corresponds
to the particular solution of case A in \S \ref{sec:basic}.  
In this particular configuration there is a balance in each ring between the
non-gravitational torque produced by the accreting flow and
the external plus internal gravitational torques 
(see Eq.~(\ref{eq:nullomega}) and Appendix A).

The halo not only alters the way in which the rings precess. 
It has been recognized for a long time that the flattening of
the dark halo is a crucial 
matter in order to understand the evolution and shape
of vertical perturbations in discs, because 
the amplitude of the warp depends on the contribution
of the dark halo to the vertical restoring force, keeping
the rings more tightly bound in a flattened halo.

In the next Sections we derive estimates of the amplitude,
warp asymmetry and evolution of the tilted disc using a formalism
that allow us to treat all the $m$-modes
in a self-consistent way, taking into account the halo flattening,
which is essential in modelling the vertical distribution. 
Our procedure is completely different to the approach followed by
LC, motivated to take a different look at the problem.
The mode decomposition will help to better understand the
underlying physics of the situation. For instance,
our calculations can tell us when
the approximation of solid rings is correct.
In the following Subsections we will discuss strictly the predictions of our 
calculations to test the viability of the accretion scenario.

\subsection{Assumptions}

The calculations start from the following assumptions:

(1) The accretion of an intergalactic flow directly onto the disc is 
the main physical agent responsible for the formation of warps. 

(2) The intergalactic flow is composed of purely baryonic matter and 
contains a negligible amount of collisionless dark matter.

(3) The flow is made of discrete collisionless clouds with a null velocity 
dispersion at infinity.

(4) The flow is plane-parallel at infinity, with the direction
and velocity constant in time.

(5) The clouds move in hyperbolic orbits, unaffected by the drag with
the hot coronal gas halo, in a galactic potential 
described by a point-mass. For the Galaxy the adopted mass
is $2\times 10^{11}$ M$_{\odot}$, which is approximately the
dynamical mass of the Galaxy within
a sphere of $16$ kpc in radius. All the mass that intersects the
disc is perfectly accreted.

(6) The inner disc and halo are frozen or non-responsive, i.e.~they
are modeled as time-independent external potentials; we do 
not include the response of the inner disc and halo to the tilt of
the disc.  The central disc and the halo are aligned.

(7) The amplitude of the warp is so small that the force caused
by the intergalactic flow does not depend on it. This assumption
together with assumption (6) imply $f_{\rm acc}(R,\phi;\theta_{0})$,
where $\theta_{0}$ is the angle of incidence of the flow.
Consequently, given $\theta_{0}$, the collisional torque exerted by the flow
is constant in magnitude and direction.

(8) The gravitational torque exerted by the flow into the disc is 
neglected. We consider only the torque produced by the collision
of the clouds with the disc. 

(9) The stellar and gas discs are gravitationally
tight enough that the differences between stellar and gas warps
are small. 

All these assumptions were also adopted by LC
except assumptions (6) and (7). We have estimated that the neglect of
the precession of the entire disc to be a reasonable approximation
(see \S\S \ref{sec:externaltorques} and \ref{sec:caveats}).
Assumption (2) is somehow related with assumption (8);  gravitational
torques by the accretion flow are expected to be small 
compared to the collisional torque
for the parameters under consideration. 
Concerning assumption (3), a null velocity dispersion of the clouds is
the simplest assumption.
From basic grounds, it can be seen that the relative amplitude of the 
$m=1$ mode respect to the $m=0$ mode depends on the velocity dispersion 
of the clouds. Regarding assumption (5), it
is likely that clouds disrupt before reaching the inner disc. For
these clouds the hydrodynamical drag with the gaseous corona cannot 
be neglected (e.g., Benjamin
\& Danly 1997). However, it is equally possible that certain dense clouds 
can achieve the far-outer parts of the disc before disruption, for a 
plausible range of parameters of the clouds. 
 
\subsection{The accretion model: Equation of motion and the mass model}
\label{sec:massmodel}
The vertical dynamics of a parcel of material of the disc  
with vertical displacement $h(R,\phi,t)$ and surface density $\Sigma (R)$, 
embedded in a dark halo with
gravitational potential $\Phi_{\rm h}$, is given by the equation
of motion
\begin{equation}
\left(\frac{\partial}{\partial t}+
\Omega \frac{\partial}{\partial \phi}\right)^{2}h=-\nu^{2}_{\rm h}h
+f_{\rm self}+f_{\rm acc},
\label{eq:verticaldyn}
\end{equation}
where $\Omega(R)$ is the circular frequency at radius $R$, 
$\nu^{2}_{\rm h}\equiv \partial^{2}\Phi_{\rm h}/\partial z^{2}|_{z=0}$ 
is the vertical frequency of the dark halo, while 
$f_{\rm self}(R,\phi,t)$ and $f_{\rm acc}(R,\phi)$ are the 
vertical forces per unit of mass on
the parcel of gas due to the self-gravity of the distorted disc ($f_{\rm self}$)
and due to the intergalactic accretion flow ($f_{\rm acc}$).

The dark halo is assumed to be frozen, i.e.~it behaves as a fixed potential,
and flattened along the symmetry axis of the disc with
axis ratio $q$, typically $q\approx 0.75$, with the classical
pseudo-isothermal density profile:
\begin{equation}
\rho_{\rm h}(R,z)=\left\{\begin{array}{ll}
\rho_{0}\frac{\exp(-a/r_{\rm t})}{1+a^{2}/r_{\rm c}^{2}} 
& \mbox{if $r\leq r_{\infty}$};\\
0 & \mbox{otherwise}, \end{array} \right. 
\end{equation}
with $a^2=R^2 +z^{2}/q^{2}$, $r_{\rm c}$ the core radius of the dark halo,
$r_{\rm t}=50 r_{\rm c}$ and $r_{\infty}=65 r_{\rm c}$, 
those last values were selected just to facilitate comparison with 
previous models, as in Binney et al.~(1998), but they do no play
any noticeable relevance in the accretion model provided they are
larger than the maximum radius of the warp $R_{2}$ (see assumption (5)). 

The warps are observed in both
the stellar disc and the gas layer and, where they can be compared
each other, the warped planes are almost the same 
(Casertano et al.~1987; Cox et al.~1996). 
Hence, within approximation (9), 
the surface density of the disc should include both stars plus gas.
We adopted a truncated exponential profile for the surface density
with scalelength $R_{\rm d}$, i.e.
$\Sigma (R)=\Sigma_{0}\exp(-R/R_{\rm d})$, whenever $R\leq R_{\rm t}$,
with $R_{\rm t}$ the truncation radius.

The vertical force $f_{\rm self}$ was given in
Eq.~(\ref{eq:selfcomplete}). Following
Binney et al.~(1998), we have adopted a softening radius
$z_{0}=0.1R_{\rm d}$.

The physical basis for the warp formation in the model of intergalactic
matter accretion 
relies on the fact that $f_{\rm acc}$ may exert a net torque in the disc 
because the flow is redirected and 
gravitationally focused by the potential of the galaxy.
With no loss of generality, we will assume that the accretion flow
comes from below the equatorial plane, i.e.~$v_{z}>0$, with a velocity 
at infinite distance of $\bmath{v}_{\infty}=(v_{x},0,v_{z})$ in 
the galaxy rest frame. We define the angle $\theta_{0}$
between the direction of the incoming gas flow at infinity
and the reference plane, i.e.~$\sin\theta_{0}=v_{z}/v_{\infty}$. 
A flow parallel to the spin axis of the galaxy has $\theta_{0}=\pi/2$.
$f_{\rm acc}$ was derived by LC in
the approximation (5). The formula for $f_{\rm acc}$ is given by 
\begin{equation}
f_{\rm acc}(R,\phi)=\frac{\rho_{\infty} v_{\infty}^{2}}{\Sigma(R)} 
\left(\frac{b}{R}\right)^{2}\frac{\partial b}{\partial R}
\frac{\partial \phi_{b}}{\partial \phi}
\sqrt{1-\cos^{2}\theta_{0}\sin^2\phi},
\label{eq:force1}
\end{equation}
where $b(R,\phi)$ is the impact parameter and $\phi_{b}(\phi)$
is the polar angle in the plane perpedicular to $\bmath{v}_{\infty}$.
The expressions for $b(R,\phi)$ and $\phi_{b}$ were given
in LC [their Eqs (38) and (34), respectively]. 

The imposed force $f_{\rm acc}$ can be expressed as a sum of
azimuthal Fourier terms of different amplitudes,
\begin{equation}
f_{\rm acc}(R,\phi)=\sum_{m=0}^{\infty} f_{m}(R)\cos m\phi.
\end{equation}
The steady response of the disc will be also a sum of Fourier modes
with amplitudes $h_{0}(R)$, $h_{1}(R)$, $h_{2}(R)$ and so on.
The rings can be treated as rigid provided that the amplitude of
the displacements for components $m=0$ and $m=1$ exceed those for all
other wavenumbers. We have checked that this is the case except for
small values of $\theta_{0}$. In fact, for $\theta_{0}<15^{\circ}$,
the $h$-amplitudes of $m=2$ mode is comparable to that for the $m=0$ component.
Hence, for $\theta_{0}>15^{\circ}$, 
we take $f_{\rm acc}(R,\phi)\simeq f_{0}(R)
+f_{1}(R)\cos\phi$ and omit the analyses for other $m$-components,
those analyses being analogous. 

The steady response of the disc has the form
\begin{equation}
h(R,\phi,t)=h_{0}(R)+h_{1}(R)\cos\left(\omega t-\phi+\phi_{0}(R)
\right),
\end{equation}
where $\omega$ is the pattern speed about the $z$-axis, 
$h_{0}(R)$ and $h_{1}(R)$ are the
amplitudes for the $m=0$ and $m=1$ modes, respectively, and $\phi_{0}(R)$ a
phase function which describes the twist of the line of maximum
heights of the pure $m=1$ mode. 
The observed integral-sign warps, or S-shape warps, 
correspond to the second term ($m=1$), whereas
the first $m=0$ mode is responsible for the generation
of the cup-shaped symmetry between the two sides of the warp.

\begin{table*}
\caption[]{Parameters for six models (columns $2$--$3$),
together with the amplitude ratio between the two sides of the
warp at $R=22 r_{\rm c}$ (column $4$) and the shape of the warp (column $5$).
The two latters were derived by solving Eqs.~(\ref{eqn:f0}) and (\ref{eqn:f1}).
S denotes integral-sign warps and L stands for one-sided warps
(see Fig.~\ref{fig:shape}). 
Type I refers to warps with increasing
amplitudes towards the edge of the disc. Asterisk indicates that
a velocity of $v_{\infty}=300$ km s$^{-1}$ instead of $100$ km s$^{-1}$
was used.}
\vspace{0.01cm}

\begin{center}
\begin{tabular}{c c c c c }\hline
{Model} & { $q$} & {$\theta_{0}$(deg)} & 
{ $\gamma_{l}$} & {Shape} \\
{}& & & {at $R=22 r_{\rm c}$}& \\

\hline
I   & 0.75   & 45  & 1.54   & S, type I \\
II   & 0.9   & 45  & 1.39   & S, type I \\
III  & 0.75  & 30  & 1.37   & S, type I \\
IV   & 0.75  & 60  & 1.97   & S, type I \\
V   & 0.75   & 75  & 5.43   & L, type I \\
VI$^{\ast}$  & 0.75   & 45  & 2.02   & S, type I \\
\noalign{\medskip\hrule\medskip}
\end{tabular}
\end{center}
\label{table:parameters}
\end{table*}

Substituting the above expressions for $h$, $f_{\rm self}$ and
$f_{\rm acc}$  into equation 
(\ref{eq:verticaldyn}) yields:
\begin{equation}
\phi_{0}(R)=0,\,\,\,\,\,\,\,\,\,\omega=0,
\label{eq:nullomega}
\end{equation}
\begin{eqnarray}
&&f_{0}(R)=\nu_{\rm h}^{2}h_{0}(R)+Gh_{0}(R)\int_{0}^{R_{\rm t}}
\Sigma (R') H(R,R') R'dR'\nonumber\\
&&-G\int_{0}^{R_{\rm t}}\Sigma(R') H(R,R')h_{0}(R')R'dR',
\label{eqn:f0}
\end{eqnarray}
\begin{eqnarray}
&&f_{1}(R)=(\nu_{\rm h}^{2}-\Omega^{2})h_{1}(R)+Gh_{1}(R)\int_{0}^{R_{\rm t}}
\Sigma (R') H(R,R') R'dR'\nonumber\\
&&-G\int_{0}^{R_{\rm t}}\Sigma(R') J(R,R')h_{1}(R')R'dR',
\label{eqn:f1}
\end{eqnarray}
with $H(R,R')$ and $J(R,R')$ as defined in Eqs.~(\ref{eq:Hmayuscula})
and (\ref{eq:Jmayuscula}), respectively.

As we anticipated in \S \ref{sec:caveats}, 
the zero pattern speed $\omega=0$, which means  
that the rings does not precess around the $z$-axis in the steady forced 
evolution, is a consequence of our assumptions (6) and (7). 
Under these approximations the form of the imposed time-independent forcing 
is $f_{\rm acc}(R,\phi)=f_{0}(R)+f_{1}(R)\cos \phi$, whose $m=1$ mode
corresponds to case A of \S\ref{sec:basic} 
(see also Appendix A).
For the same reason, the twist function vanishes, $\phi_{0}=0$, implying that
the line of maximum heights is straight; the direction of the
torque produced by the accreting flow provides a direction along which
the warp's line of node may naturally align. 
Unfortunately, the predicted value $\omega=0$ is not in accord with
the pattern speed inferred from the stellar velocity distribution
in the warped Galactic disc of $\omega < -20$ km s$^{-1}$kpc$^{-1}$ 
(Smart et al.~1998).

The integral equations (\ref{eqn:f0}) and (\ref{eqn:f1}) can be solved 
by treating the continuous disc as a collection of massive
rings and using standard methods.
We will use units such as $G=r_{\rm c}=1$ and the halo mass is 
unity. For the disc mass we choose $0.1$. Note that our mass model
is identical to that adopted by Binney et al.~(1998).
The disc is made up by $60$ rings uniformly spaced in radius
between $R=9 r_{\rm c}$ and $R=R_{\rm t}=22 r_{\rm c}$.
The results for four representative models with different angles 
of the wind and flattenings of
the dark halo, as described in Table \ref{table:parameters}, will be explored 
and used as a test for the accretion model in the next Subsection.

\subsection{Amplitude and shape of the warps: Results, 
interpretation and discussion}
\label{sec:test}
It is convenient to start our discussion by considering the baryon
density and flow velocity required to explain the observed amplitudes
of galactic warps.
Later on, we will consider the observable asymmetries thereby detectable
in H\,{\sc i} velocity contour maps and in
edge-on galaxies, as a result of the coexistence of the modes $m=0$ and 
$m=1$.  
A comparison of predictions with observations will be used to validate 
or rule out the accretion model. 

\subsubsection{Amplitude}
\label{sec:amplitude}
\begin{figure*}
\epsfig{file=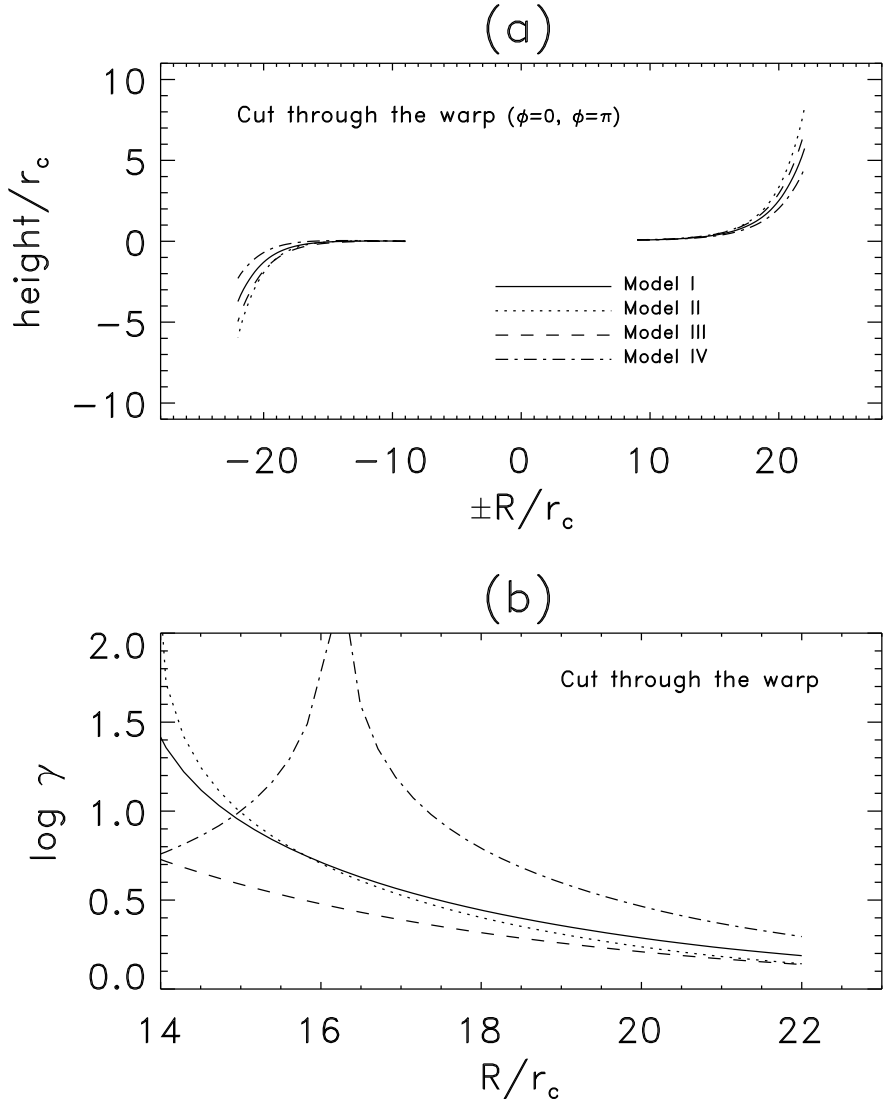,
        angle=0,
        width=5.3in
}
\caption{Shape of the warp along a cut through the (straight) line
of maximum heights for the different models of Table \ref{table:parameters}
(panel {\bf a}) The
radial dependence of $\gamma$, which is a measure of the warp
amplitude asymmetry is shown in panel {\bf b}. 
For making estimates in the context of the Galaxy, $r_{c}\simeq 0.7$ kpc.}
\label{fig:shape}
\end{figure*}

Figure \ref{fig:shape}a displays the height of the disc respect to the
symmetry plane of the halo versus the radius of the disc along a
cross-section cut through the warp for some models specified in Table
\ref{table:parameters}. In all the models, 
we have used the same parameters for the flow than in LC,
$v_{\infty}=100$ km s$^{-1}$ and 
$\rho_{\infty}= 6\times 10^{-5}$ cm$^{-3}$. This density is likely
a generous assumption (see \S \ref{sec:caveats}). 
For a galaxy like the Milky Way, $R_{\rm d}=4.3r_{c}=3$ kpc. 
It turns out that the observed Galactic warp amplitude can be reproduced with
the adopted parameters.
We confirm the results of LC derived in a very different framework.

For cluster galaxies, both $\rho_{\infty}$
and $v_{\infty}$ can be significantly larger than the values quoted.
Nevertheless the predictions of  
the accretion scenario are not longer valid for these galaxies because
the assumptions behind the accretion model are
not a fair representation of the collisional interaction
between the intracluster medium and the interstellar gas (e.g.,
S\'anchez-Salcedo 2004). Numerical calculations of the ram pressure
stripping process are required to explain the distorted velocity field and
extraplanar (usually resembling a U-rim) 
distribution of the gas observed in these galaxies
(e.g., Vollmer et al.~2000). 
Since the problem of the persistence of warps is circunscribed to relatively
isolated galaxies, we will omit any discussion regarding the origin
of warped galaxies in clusters.

Fixed $\rho_{\infty}$ and $v_{\infty}$ at the preferred values, 
a scaling relation between
the size of the galaxy, $R_{l}$, defined as the last H\,{\sc i}
measured point and warp's amplitude could be expected.  
The accretion force given in Eq.~(\ref{eq:force1}) depends on
the parameter $GM/Rv_{\infty}^{2}$. To a good approximation and for
the range of interest, it can be fitted by $f_{\rm acc}\propto 
(GM/Rv_{\infty}^{2})^{1.3}\simeq (v_{\rm rot}/v_{\infty})^{2.6}$,
where we have used that $M$ is the dynamical mass within $R_{l}$,
as adopted in previous Sections. The displacement $h$ at the end
of the warp $h_{l}$ scales according to $f_{\rm acc}\approx 
h_{l}v_{\rm rot}^{2}R_{l}^{-2}$ (see Eq.~\ref{eq:gunngott}).
Therefore $h_{l}\sim v_{\rm rot}^{0.6} R_{l}^{2}$ and we would
expect a positive correlation between the amplitude of the
warp and two parameters: galaxy radius and galaxy mass.
In contrast, Castro-Rodr\'{\i}guez et al.~(2002) found a slight
anticorrelation between the observed amplitude and these parameters.
In fact, the amplitude of warps in dwarf galaxies cannot be accounted
by the accretion model itself.

\subsubsection{Shape and morphological asymmetries}

\begin{figure}
\epsfig{file=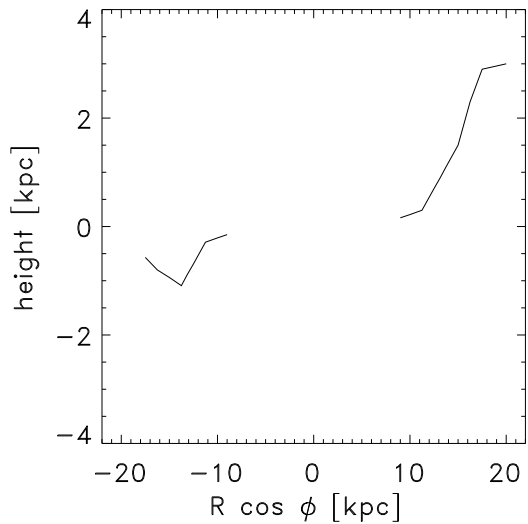,
        angle=0,
        width=3.3in
}
\caption{Shape of the warp of the Milky Way taken from
Burton (1988) with $R_{\odot}=8$ kpc.}
\label{fig:milky}
\end{figure}

As it becomes apparent in panel (a) of Fig.~\ref{fig:shape},
the accretion model predicts that all the S-warps must be of type I
in the terminology of Sparke \& Casertano (1988), since the inclination of the
outer warp increases from the center to the edge of the disc.
This result is mainly a consequence of the dependence of the force
term with the surface density $f_{\rm acc}\propto \Sigma^{-1}$ 
(see Eq.~(\ref{eq:force1})).
If the flow is smooth and accretion is the only cause of warps,
it is very difficult to explain why it is very
common to find that one side turns back up toward the inner-disc
plane after reaching its maximum vertical height (Burton 1988;
Garc\'{\i}a-Ruiz et al.~2002), implying that the amplitude of
the $m=0$ component relative to that of the $m=1$ mode increases
with galactocentric radius. As an example,
the curve of the asymmetric warp of the Milky Way is shown in
Fig.~\ref{fig:milky} from Burton (1988)
with $R_{\odot}=8$ kpc. For our Galaxy, the ratio between
the amplitude of the mode $m=0$ over the amplitude of mode $m=1$
increases from $\sim 0.1$ at $14$ kpc to $0.7$ at $18$ kpc.
This radial behaviour is unlikely in the accretion model as
the $m=1$ response becomes dominant at the outer parts due
to the near equality of the orbital and the vertical oscillation
periods at successively larger radii. As it is illustrated
in Fig.~\ref{fig:shape}a, the warps never curve back to the disc-plane. 

In the accretion model warps should be asymmetric (see
panel (a) of Fig.~\ref{fig:shape}). Given the mass model
of a certain galaxy, the asymmetry of the warp depends on
$\theta_{0}$, $v_{\infty}$, $q$, and the velocity dispersion of the clouds,
which was assumed to be zero.
Smaller the angle $\theta_{0}$ is, more antisymmetric (i.e., they are
closer to a integral-sign shape) the warp becomes,
but even for $\theta_{0}=30^{\circ}$, the amplitudes between the two
sides in the warp are significantly different (see Table 
\ref{table:parameters}). It is interesting to note that 
for the model with $\theta_{0}=45^{\circ}$, the ratio between 
the amplitudes $f_{1}/f_{0}\approx 1$, fairly independent of $R$ in
the range of interest, but the disc is more
responsive to the $m=1$ mode, resulting in $h_{1}/h_{0}\approx 5$ at
$22r_{c}$. The reason is that the $m=0$ amplitude $h_{0}\propto
\nu^{-2}_{t}$ (in the notation of \S \ref{sec:basic}), 
whereas $h_{1}\propto (\nu^{2}_{t}-\Omega^{2})^{-1}$.
In panel (b) of Fig.~\ref{fig:shape} we have plotted the radial
dependence of the value $\gamma(R)$, defined as the ratio between
the amplitudes at the same $R$, as measured from the central
galaxy plane to the peak of H\,{\sc i} column density, on both sides
of the warp with the convention that
$\gamma(R) \geq 1$. For model IV,  
$\gamma\rightarrow\infty$ at a certain radius near $R\simeq 16 r_{\rm c}$.
This divergence is because at that radius 
the warp crosses the equatorial plane of the inner disc
at one side. For angles $\theta_{0}>75^{\circ}$ warps bent with 
a L-shape (see model V in Table \ref{table:parameters}) 
or even a bowl-shape. Hence, 
we expect a percentage of occurrence for these types of
warps of $\sim 3.5$\%. This fraction may be larger if the clouds
have no-null velocity dispersion at infinity. 
In fact, if the flow velocity has its origin in the motion of the galaxies, 
one expects similar peculiar velocities for the clouds.
In that case, i.e.~clouds having an intrinsic velocity comparable to
$v_{\infty}$, the percentage of bowl-shaped
warps increases by a factor $\sim 2$.

Observations show that the integral-sign 
warps are not perfectly symmetric (Sancisi 1976; Burton 1988; Bosma 1991; 
Garc\'{\i}a-Ruiz et al.~2002). 
The natural question that arises is whether the accretion model can
explain the degree of asymmetry of warps of type I.
To check this possibility, we have selected from
the literature a sample of $16$ galaxies with type I warps
and estimated $\gamma_{l}$, the value of $\gamma$ 
at the end of the warp.
We have not included those galaxies that are strongly perturbed
by the interaction with nearby neighbours or with very small warps. 
The selected galaxies are: M33 (Reakes \& Newton 1978; Corbelli 
\& Schneider 1997), NGC 4244 (Sancisi 1976), NGC 300
(Rogstad et al.~1979), NGC 2841 (Begeman 1987),
NGC 4013 (Bottema 1996), UGC 7170 (Cox et al.~1996),
ESO 123-G23 (Gentile et al.~2002), NGC 5055 (Battaglia et al.~2005),
U6964, U7774 and U8711 from Garc\'{\i}a-Ruiz et al.~(2002) and
NGC 2685, NGC 3718, NGC 5204, UGC 3580 and NGC 2541 (J\'{o}zsa et al.~2004
and private communication). 
All the galaxies of this sample present 
integral-sign shaped warps with $\gamma_{l} <2.15$, and most of them
are so symmetric that no difference between the amplitude of both
sides has been reported. 
In order to increase statistical significance, we have
considered the distribution of the $\gamma_{l}$-values for the
sample of $137$ warped galaxies observed in the 
optical band (Castro-Rodr\'{\i}guez
et al.~2002). We find that only $36\%$ of these galaxies 
present optical warps with $\gamma_{l}>1.25$. 
Note that since stars
extend less than the gas layer, the stellar warp is expected to
be even more asymmetrical (see Fig.~\ref{fig:shape}b). 
The existence of galaxies with small values of $\gamma_{l}$ is difficult
to explain if accretion is the main and only responsible for warps.

From studies of the asymmetric shape of the warps, 
we conclude that accretion flows may play a role in warping
the discs, but they cannot be the only cause of warps.

\subsection{The vertical distribution in H\,{\sc i} volume density}
S\'anchez-Salcedo (2004) pointed out that if the force responsible
for the warp has a hydrodynamical origin,
rather than gravitational as in the cosmic infall scenario, the
vertically shifted disc must be supported by a pressure gradient.
The reason is that one can impose the condition of
hydrostatic equilibrium
as the sound crossing time over the height of the disc at the sound
speed of $10$ km s$^{-1}$, is much shorter than the period for the
gas to cross two successive crests of the warp. Consequently, the volume
density of gas in a cut along the vertical axis $z$, at a given $(R,\phi)$,
must present a very asymmetric vertical distribution, but this seems to be
in conflict with the observed fairly symmetric distribution
of the Galactic warp by Diplas \& Savage (1991).
In addition, a high rate of energy deposition
in the cold disc by the ram pressure might produce other observable
consequences, such as the ionization of the H\,{\sc i} layer (Kenney \&
Koopmann 1999).
 
\subsection{The thickness of the outer H\,{\sc i} disc}
The scaleheight of the outer disc may provide additional
constraints on the strength of the ram pressure acting on discs.
In addition to the asymmetric distribution in a cut
along the vertical $z$-axis discussed in the previous Section, 
the external dynamical pressure caused by falling clouds,
will produce a vertical compression of the H\,{\sc i} disc. This
compression may have important observational consequences for the outer
thickness of the H\,{\sc i} disc, such as a reduction and
a large-scale asymmetry in the H\,{\sc i} scaleheight.
The asymmetry is expected as the result of the azimuthal dependence 
of this external pressure, which ultimately is responsible for the S-warp. 
The recent work by Narayan et al.~(2005) implies there is 
little room for asymmetric infall of material at the outer warp.  
In fact, we suggest that studies of the H\,{\sc i} scaleheight 
can place strong constraints on the accretion rate of intergalactic gas.
To illustrate this point, we are going to estimate the thickness of
the far-outer regions of the Galactic disc
for the preferred parameters of the accretion model: 
$\theta_{0}=45^{\circ}$, $\rho_{\infty}=6\times 10^{-5}$ cm$^{-3}$
and $v_{\infty}=100$ km s$^{-1}$. 
When comparing with observations we will adopt a value $R_{\odot}=8$ kpc.

The external pressure due to accretion 
can be written as $\alpha(R,\phi)\rho_{\infty}v_{\infty}^{2}$, where
the dimensionless function $\alpha(R,\phi)$ can be derived from
Eq.~(\ref{eq:force1}).  At a galactocentric distance
of $22.5$ kpc $\alpha\approx 40$ at $\phi=0$, and declines
to a value $\sim (40/6)$ at the opposite side $\phi=\pi$.
An upper limit on the thickness of the disc, 
as measured as the height containing 
$63\%$ of the H\,{\sc i} surface density, based on the equation from
equilibrium is:
\begin{equation}
L\leq{\rm min}\left[\frac{2 c_{g}^{2}}{\pi G (\Sigma_{g}+
(c_{g}/c_{\star})\Sigma_{\star})},
\frac{\Sigma_{g}c_{g}^{2}}{\alpha\rho_{\infty}v_{\infty}^{2}}\right],
\label{eq:thickness}
\end{equation} 
where $\Sigma_{g}$ and $\Sigma_{\star}$ are the surface density
of gas and stars, respectively, $c_{g}$ is the effective gaseous
vertical velocity dispersion, including cosmic rays and magnetic
fields, and $c_{\star}$ the stellar velocity dispersion.
The first expression inside the parenthesis corresponds to the 
thickness of an isothermal self-gravitating sheet in the lack of 
any ambient pressure (e.g., Elmegreen 2002), while the second
expression gives the characteristic thickness of an isothermal 
gas layer that is pushed in a cylinder from one side by a piston
moving with constant acceleration, neglecting
self-gravity. The action of the piston mimics the pressure of a
rain of clouds falling on one cap of the disc.
Since both self-gravity and dynamical pressure
occur simultaneously in the accretion model, Eq.~(\ref{eq:thickness}) 
necessarily provides 
a boundary for $L$. To evaluate $L$, we need the H\,{\sc i}
velocity dispersion, which is observed to be almost constant with
radius and is about $9\pm 1$ km s$^{-1}$ in the inner Galaxy
(Spitzer 1978; Malhotra 1995). Beyond the solar circle, however,
the dispersion is not yet measured. Studies of external galaxies
show that the observed velocity dispersion has a narrow range
(e.g., Lewis 1984; Shostak \& van der Kruit 1984; Dickey et al.~1990).
Aiming at conservative estiamtes, we will take a dispersion 
of $10$ km s$^{-1}$. If $c_{g}\sim 7$ km s$^{-1}$
without contribution from magnetic effects (as measured directly
by the H\,{\sc i} velocity dispersion in some galaxies), the 
scaleheights would be about a factor $2$ smaller.
For $\Sigma_{g}=1$ M$_{\odot}$ pc$^{-2}$ and $\Sigma_{\star}\approx 0$ 
at $R=22.5$ kpc (Wouterloot et al.~1990)
and $c_{g}=10$ km s$^{-1}$, we get $L<180$ pc at $\phi=0$ and
$L<1100$ pc at $\phi=\pi$, and half of that for $c_{g}=7$ km s$^{-1}$.   
This variation of at least a factor of $\sim 6$ in azimuth, is not
supported by observations that show very little net asymmetry in the
derived thickness of the H\,{\sc i} layer (Henderson et al.~1982;
Merrifield 1992). Moreover, a thickness of $90$--$180$ pc at $\sim 22.5$ kpc
is in serious conflict with the observed measurements of $~2200$ pc
in the Galaxy (Wouterloot et al.~1990) and of $\sim 1$ kpc
in external galaxies (e.g., Sancisi \& Allen 1979;
Brinks \& Burton 1984; Olling 1996). We find that
it is not possible to bend the outer disc a kpc-height without
making the scaleheight too small.
By imposing that the outer disc is not appreciably compressed by 
the cloud rain, the height $h$ derived in \S \ref{sec:test} should
be reduced by a factor $\sim 10$.

\section{Conclusions}
The failed of the theory of bending modes of self-gravitating
discs in explaining the persistence of galactic warps (Nelson \&
Tremaine 1995; Binney et al.~1998) has led to invoke external torques
caused by the interaction of the galaxy discs with their environment.
The intergalactic magnetic field and the accretion of a teneous
intergalactic gas onto the disc have been suggested as possible
causes for galaxies to be warped. The fact that galaxies in poor
environments are even more frequently warped than galaxies in
dense enviroments (Garc\'{\i}a-Ruiz et al.~2002) and other 
empirical correlations (Battaner et al.~1991; Castro-Rodr\'{\i}guez
et al.~2002; S\'{a}nchez-Saavedra et al.~2003) make the above 
formation mechanisms very attractive and it seems appropriate
to seek ways in which we can test their viability.

We have shown that magnetically induced warps suffer from 
an acute winding problem as differential rotation is unimpeded
under frozen field conditions and hence warps cannot persist.
Curiously, the time-scale for winding of the warp is shorter than
in the non-magnetic case studied by Kahn \& Woltjer (1959).
It remains unclear why the different warps
of neighbouring galaxies should be aligned if the magnetic
fields are dragged by galactic circular rotation. 
We have considered the role of the inclusion of turbulent magnetic
diffusion and the $\alpha$-effect. If the magnetic turbulent
diffusivity is not quenched, the amplitude of the warp diminishes
in one local diffusivity time-scale $\sim H^2/\eta_{t}$, which
is $\approx 0.3$ Gyr for $\eta_{t}\approx 10^{26}$ cm$^{2}$ s$^{-1}$.  
Moreover, for galaxies moving through the intergalactic medium, 
the plane-parallel
configuration of intergalactic magnetic fields around galaxies,
as required in the magnetic model, is by no means expected because
of the shear terms.

Accretion of infalling diffuse material from the intergalactic
medium may be inferred from current low-mass star formation rates
in spiral galaxies. In the model of cosmic infall, the new accreted
material reorients the surrounding halo potential, as
skew angular momentum is absorbed, and antisymmetric integral-sign warps can be
excited even if the accreted matter was purely dissipative gas
(e.g., Jiang \& Binney 1999; Bournaud \& Combes 2002). 
The situation is completely
different in the case of oblique accretion onto the disc 
of an intergalactic flow caused by the relative motion of the 
galaxies through the IGM. In such scenario, 
the flow has no net angular momentum at infinite distance but, 
when it reaches the disc,
the impulse transfer has an azimuthal dependence and both the $m=0$ and 
the $m=1$ modes are excited, generating asymmetric warps. 
We have inferred the amplitude and level of warp asymmetry in the
accretion model following a completely different formulation than
in LC. It turns out that, 
as reported by LC, an accretion flow velocity of $100$ km s$^{-1}$ and
densities of $6\times 10^{-5}$ cm$^{-3}$ can account for the amplitude
of the Galactic warp. 
However, the model presents difficulties in explaining the existence
of fairly antisymmetric warps and, hence, accretion of a plane-parallel
flow cannot be the only cause of warps. 
For the Galactic model, the warp should
be of type I with the maximum gas displacements on the two sides
in the ratio $3:2$. In addition, it predicts that a 
fraction $\sim 3.5$\% of warped galaxies, at least, should be of type L or U; 
this fraction may be larger if the clouds have a no-null velocity dispersion.
There are other predictions that are not borne out by
the observations: (1) this model predicts that antisymmetric
warps should be of type I, whereas those warps falling back towards
the midplane of the inner disc should present a U-rim shape;
(2) if accretion is the only mechanism to excite warps, 
a strong correlation between amplitude of the warp and size
of the H\,{\sc i} disc is expected,
(3) the rather symmetric vertical distribution of the volume density in
the warp of our Galaxy indicates that the pressure by the action of
the clouds rain on one cap of the H\,{\sc i} disc 
does not produce an appreciable vertical distortion; 
(4) the external pressure required to bend the outer disc would produce 
H\,{\sc i} scaleheights too small; (5) the thickness of the H\,{\sc i}
disc should have a strong dependence on azimuth.
We suggest that the dynamical pressure needed to take the gas off the
plane is a factor $\sim 10$ much larger than observations permit.

\section*{acknowledgements}
I am grateful to M.~L\'{o}pez Corredoira for helpful comments
and intense exchange of arguments. A.~Hidalgo G\'amez and E.~Moreno are thanked 
for commenting on the first draft of this manuscript. 
I thank an anonymous referee for pointing out the relevance of
the turbulent magnetic diffusion.
This work was supported by CONACYT project 2002-C40366.

\appendix
\section{Analysing the dynamics of a disc under a torque of type A}
\subsection{Response of the disc and temporal evolution}

We start examining the motion of a disc cloud in Cartesian
coordinates $(x_{c},y_{c},z_{c})$, originally
located in the $xy$ plane (i.e.~$z_{c}=\dot{z}_{c}=0$ at $t=0$),
under the action of an axisymmetric gravitational potential
created by the disc and the dark halo,
$\Phi(R,z)$, and a vertical
nongravitational force $f_{\rm ng}(R,\phi)=f_{1}(R)\sin\phi$
as described in case A of \S \ref{sec:basic}. 
Let us consider only displacements of the cloud along
the $z$ axis. The equation of motion for the three 
coordinates are:
\begin{equation}
\ddot{x}_{c}=-\Omega^{2}x_{c},
\end{equation}
\begin{equation}
\ddot{y}_{c}=-\Omega^{2}y_{c},
\label{eq:ycI}
\end{equation}
\begin{equation}
\ddot{z}_{c}=-\nu_{t}^{2}z_{c}+f_{1}(R)\sin\phi.
\label{eq:zcI}
\end{equation}
Integrating the equations for $x_{c}$ and $y_{c}$, the azimuthal
angle of a cloud evolves according to $\phi=\phi_{0}+\Omega t$, where
$\phi_{0}$ is the azimuthal angle of the cloud at $t=0$. Hence
$y_{c}(t;\phi_{0})=R\sin(\Omega t+\phi_{0})$ and Eq.~(\ref{eq:zcI})
becomes
\begin{equation}
\ddot{z}_{c}=-\nu_{t}^{2}z_{c}+f_{1}\sin(\Omega t+\phi_{0}).
\end{equation}
The solution of this equation after imposing the assumed vertical
displacement and speed at $t=0$ is
\begin{equation}
z_{c}(t;\phi_{0})=z_{I}\sin(\Omega t+\phi_{0})-z_{H}\sin(\nu_{t} t+\phi_{H}),
\end{equation}
where 
\begin{equation}
z_{I}=\frac{f_{1}}{\nu_{t}^{2}-\Omega^{2}},
\end{equation}
\begin{equation}
z_{H}(\phi_{0})=z_{I}\frac{\Omega}{\nu_{t}}(\cos\phi_{0})
\left[1+\frac{\nu_{t}^{2}}{\Omega^{2}}
\arctan^{2}\phi_{0}\right]^{1/2},
\end{equation}
and
\begin{equation}
\phi_{H}(\phi_{0})=\arctan\left(\frac{\nu_{t}}{\Omega}\tan\phi_{0}\right).
\end{equation}
As discussed at length in \S \ref{sec:basic}, 
the component $z_{H}\sin(\nu_{t} t+\phi_{H})$
is erased by the winding process and hence, after a few dynamical
time-scales, only the stationary configuration 
\begin{equation}
z_{c}(t;\phi_{0})=\frac{f_{1}}{\nu_{t}^{2}-\Omega^{2}}\sin(\Omega t+\phi_{0}),
\end{equation}
is observable. In this particular configuration, a cloud oscillates
with frequency $\Omega$ in the vertical direction. 
The vertical displacement $h(R,\phi,t)$ can be immediately
calculated from $z_{c}(t;\phi_{0})$ by using the identity
$\phi_{0}=\phi-\Omega t$, i.e.~$h(R,\phi,t)=z_{c}(t;\phi-\Omega t)$.

\subsection{Balance of torques in the stationary configuration} 
In \S \ref{sec:basic} it was shown that
\begin{equation}
h(R,\phi,t)=\frac{f_{1}}{\nu_{t}^{2}-\Omega^{2}}\sin \phi,
\label{eq:secondtime}
\end{equation}
is the only configuration that describes a stationary
warped disc in case A. We wish to check that the total torque
acting on each ring is zero.

The total (gravitational plus non-gravitational)
force per unit mass acting on a parcel of gas at the 
position $(x,y,z)$ is
\begin{equation}
\bmath{f}=-\Omega^{2}x\bmath{e}_{x}
-\Omega^{2} y\bmath{e}_{y}+(f_{1}\sin\phi-\nu_{t}^{2}z)\bmath{e}_{z}.
\end{equation}
The torque associated with this force is
\begin{eqnarray}
&&\bmath{\tau}=\bmath{r}\times \bmath{f}=
\left[(\Omega^{2}-\nu_{t}^{2})y z+y f_{1}(R)\sin\phi\right]\bmath{e}_{x}\nonumber\\
&&+\left[(\nu_{t}^{2}-\Omega^{2})x z-x f_{1}(R)\sin\phi\right]\bmath{e}_{y}.
\label{eq:torquet}
\end{eqnarray} 
The torque acting on a ring of radius $R$ is obtained by integrating
over the azimuthal angle
\begin{equation}
\bmath{\tau}_{\rm ring}(R)=\int_{0}^{2\pi} \bmath{\tau} d\phi.
\label{eq:torquetring}
\end{equation}
In cylindrical coordinates we know that
$x=R\cos\phi$ and $y=R\sin\phi$. From Eq.~(\ref{eq:secondtime}) it 
holds that $z=(\nu_{t}^{2}-\Omega^{2})^{-1}f_{1}\sin\phi$.
Substituting these expressions into 
Eqs~(\ref{eq:torquet})-(\ref{eq:torquetring}) and performing
the $\phi$-integration we obtain $\bmath{\tau}_{\rm ring}(R)=0$, 
implying that the total (disc$+$halo$+$non-gravitational torque)
torque in a ring vanishes, which is fully consistent
with the fact that the rings do not precess.

\section{Evolution of the magnetic field in a rotating disc
with vertical motions}
\subsection{Induction equation and magnetic evolution in a
imposed velocity field}
Vertical motions associated with the vertical displacement of
the disc and differential rotation
can both modify the configuration of the magnetic lines in the
disc. In this appendix we consider the evolution of the magnetic field
under field freezing conditions in a steady tilted disc.
We start with the full equations for 
$\partial B_{R}/\partial t$, $\partial B_{\phi}/\partial t$
and $\partial B_{z}/\partial t$ that result from the induction 
equation (\ref{eq:induction}) by a imposed velocity field of the form
\begin{equation}
\bmath{v}=v_{\phi}(R)\bmath{\hat{e}}_{\phi}+v_{z}(R,\phi)\bmath{\hat{e}}_{z}.
\end{equation}
In the absence of turbulent motions, the resulting equations are
\begin{equation}
\frac{DB_{R}}{Dt}=0,
\label{eq:app1}
\end{equation}
\begin{equation}
\frac{DB_{\phi}}{Dt}=R\frac{d\Omega}{dR}B_{R},
\label{eq:app2}
\end{equation}
\begin{equation}
\frac{DB_{z}}{Dt}=B_{R}\frac{\partial v_{z}}{\partial R}+
\frac{B_{\phi}}{R}\frac{\partial v_{z}}{\partial \phi},
\label{eq:app3}
\end{equation}
where $D/Dt\equiv (\partial/\partial t+\Omega\partial/\partial \phi
+v_{z}\partial/\partial z)$. In the derivation of these equations
we have adopted the simplifying assumption that 
$\partial B_{z}/\partial z\approx 0$ across the thickness of the disc.
Since the magnetic force is given by the term $B_{R}\partial B_{z}/\partial R$,
we are mainly interested in the evolution of $B_R$ and $B_{z}$.
According to Eq.~(\ref{eq:app1}), $B_{R}$ is constant in an element
volume moving
with the fluid. However, vertical motions could lead
to variations of $B_{z}$ due to the gradients $\partial v_{z}/\partial R$
and $\partial v_{z}/\partial \phi$.
  
In order to estimate the importance of these terms, let us assume
that the dynamics of the disc can be described by 
$h(R,\phi,t)=h_{1}(R)\sin(\phi-\omega t)$, where $\omega(R)$
is a known azimuthal frequency which may depend on $R$.
With the help of the definition of $v_{z}$ given in 
Eq.~(\ref{eq:defvz}), substituting 
into Eq.~(\ref{eq:app3}), and exploiting the relations (\ref{eq:app1}) and
(\ref{eq:app2}), $B_{z}(R,\phi,t)$ can be obtained after integration.
Define the initial magnetic field $\bmath{B}^{(0)}$ by
\begin{equation}
\bmath{B}^{(0)}(R,\phi)=\bmath{B}(R,\phi,0),
\end{equation}
then $B_{z}$ is given by the sum of three terms
\begin{equation}
B_{z}(R, \phi,t)=B_{z}^{(0)}(R,\phi-\Omega t)+
B_{z}^{(a)}(R,\phi,t)+B_{z}^{(b)}(R,\phi,t)+B_{z}^{(c)}(R,\phi,t),
\end{equation}
where
\begin{equation}
B_{z}^{(a)}(R,\phi,t)=B_{R}^{(0)}(R,\phi-\Omega t)\frac{dh_{1}}{dR}
\left[\sin\left(\phi-\omega t\right)-\sin\left(\phi-\Omega t\right)\right],
\label{eq:compa}
\end{equation}
\begin{equation}
B_{z}^{(b)}(R,\phi,t)=B_{\phi}^{(0)}(R,\phi-\Omega t)\frac{h_{1}}{R}
\left[\cos\left(\phi-\omega t\right)-\cos\left(\phi-\Omega t\right)\right],
\label{eq:compb}
\end{equation}
and
\begin{equation}
B_{z}^{(c)}(R,\phi,t)=B_{R}^{(0)}(R,\phi-\Omega t)h_{1}
\frac{d(\Omega-\omega)}{dR}t\cos\left(\phi-\omega t\right).
\label{eq:compc}
\end{equation}
Although this solution is mathematically exact, the growth
of magnetic gradients will perturb the fluid motions
$\bmath{v}$ and, therefore, this solution is only valid at times short
enough that one can ignore the back reaction.

The first special case occurs when $\omega(R)=\Omega(R)$. Under this
condition $B_{z}^{(a)}=B_{z}^{(b)}=B_{z}^{(c)}=0$ and therefore
$B_{z}(R,\phi,t)=B_{z}^{(0)}(R,\phi-\Omega t)$. The question arises as
to how $\partial B_{z}/\partial R$ varies. To do so, one has to
remind that 
\begin{equation}
\frac{\partial \bmath{B}^{(0)}(R,\phi-\Omega t)}{\partial R}=
\frac{\partial \bmath{B}^{(0)}(R,\phi)}{\partial R}-
\frac{\partial \bmath{B}^{(0)}(R,\phi)}{\partial \phi} 
\frac{d\Omega}{dR} t.
\label{eq:movingderivative}
\end{equation}
Therefore, if initially $B_{z}$ depends on the azimuthal variable $\phi$,
then $\partial B_{z}/\partial R$ grows linearly in time due to differential
rotation. $\partial B_{z}/\partial R$ will be time-independent only
if $\omega=\Omega$ and either $B_{z}$ is axisymmetric or the disc
rotates in a solid-body manner (or both).
Summing up, in a differentially rotating disc with $\omega=\Omega$ and
$B_{z}$ being a function only of $R$, the forcing term
$B_{R}\partial B_{z}/\partial R$ is constant in an element volume
moving with the fluid. In the following we discuss the case 
$\omega\neq \Omega$.

In the magnetic scenario for warps formation, we know from
Eq.~(\ref{eq:initialfield}) that both
$B_{R}^{(0)}$ and $B_{\phi}^{(0)}$ depend on the azimuthal angle $\phi$.
Thus, the radial derivatives of $B_{z}^{(a)}$, $B_{z}^{(b)}$ 
and $B_{z}^{(c)}$ will contain a term $\propto td\Omega/dR$ due to
the second term of the RHS of Eq.~(\ref{eq:movingderivative}). 
Hence, the magnetic gradient $\partial B_{z}/\partial R$
will grow in time and, therefore, no steady-state configuration exists
in a differential rotation disc.

In addition, from Eqs.~(\ref{eq:compa})--(\ref{eq:compc}) it follows that,
in the presence of vertical motions\footnote{Remind
that $v_{z}=(\partial/\partial t+\Omega\partial/\partial\phi)h$.
Consequently, $v_{z}\neq 0$ implies $\omega\neq\Omega$.} 
($\omega\neq\Omega$),
$B_{z}$ and hence $\partial B_{z}/\partial R$ depend on $\phi$,
even if $B_{z}$ were initially independent of $\phi$.
Therefore, the magnetic term $B_{R}\partial B_{z}/\partial R$,
which was supposed initially of the form $\sin(\phi+\zeta)$,
with $\zeta$ a constant phase, 
as required to drive integral-sign warps (see \S 3),
will develop undesired modes with $m> 1$, for example, 
$\propto \sin(\phi+\zeta) \sin(\phi+\xi)$ and so on. These modes of higher
wavenumber would spoil the original $m=1$ shape of the warp.
In the next Subsection we will estimate the evolutionary time-scales
for the growth of magnetic gradients.

Finally, we must remark that if $\omega\neq \Omega$ then $B_{z}(R,\phi,t)$ 
will depend on $\phi$.
Hence, the term $(\hat{\rho} R)^{-1}B_{\phi}\partial B_{z}/\partial \phi$
that appears in Eq.~(\ref{eq:magtorque}) increases with time.

\subsection{Evaluation of the magnetic evolutionary time-scales in a 
galactic disc}
As discussed in the
previous Subsection, in order to preserve the ability of the magnetic torque
to maintain a $m=1$ distortion during a time $t_{d}$, 
the magnetic gradients should change only slightly in that time.
We introduce the characteristic growth time-scales of the magnetic
stress component $B_{R}\partial B_{z}/\partial R$, $\tau^{(n)}$
with $n=a,b,c$, as the times that are required for 
$\partial B_{z}^{(n)}/\partial R$
to change by of order of $\partial B_{z}^{(0)}/\partial R$ itself, 
according to the following relation
\begin{equation}
\frac{\partial B_{z}^{(n)}}{\partial R}\bigg|_{\tau^{(n)}}=
\frac{\partial B_{z}^{(0)}}{\partial R}.
\end{equation}
We are now in a position to estimate these characteristic time-scales
by performing the derivatives with respect to $R$
in Eqs.~(\ref{eq:compa})--(\ref{eq:compc}). 

Given $\chi (R)$ a certain variable that depends on $R$, it is convenient
for notation to define $\Delta \chi \equiv |\chi(R_{2})-\chi(R_{1})|$,
where $R_{1}$ and $R_{2}$ are the minimum and maximum radii of the warp.
For example, $\Delta h_{1}=|h_{1}(R_{2})-h_{1}(R_{1})|=h_{1}(R_{2})$ 
is the height of the warp at $R_{2}$. With this notation, the tip
warp angle denoted by $\theta_{w}$ is given by $\tan\theta_{w}\equiv
h_{1}(R_{2})/(R_{2}-R_{1})=\Delta h_{1}/\Delta R$.

The characteristic growth time-scales for the different contributions are
\begin{equation}
\tau^{(a)}={\rm max}\{\frac{\Delta B_{z}^{(0)}}{\bar{B}_{R}^{(0)}}
\frac{\Delta R}{\Delta h_{1}}\frac{1}{\Delta\Omega},\frac{\pi}{2|\Omega
-\omega|}\},
\end{equation}
where the bar denotes a characteristic value in the warp, say
$\bar{\chi}\equiv  \chi([R_{2}+R_{1}]/2)$.
Analogously,
\begin{equation}
\tau^{(b)}={\rm max}\{\frac{\Delta B_{z}^{(0)}}{\bar{B}_{\phi}^{(0)}}
\frac{\Delta R}{\Delta h_{1}}\frac{1}{\Delta\Omega},\frac{\pi}{2|\Omega
-\omega|}\},
\end{equation}
and
\begin{equation}
\tau^{(c)}={\rm min}\{\tau_{1},\tau_{2},\tau_{3} \},
\end{equation}
with
\begin{equation}
\tau_{1}=\frac{\Delta B_{z}^{(0)}}{\Delta B_{R}^{(0)}}
\frac{\Delta R}{\Delta h_{1}}\frac{1}{\Delta\omega'},
\end{equation}
\begin{equation}
\tau_{2}=\left(\frac{\Delta B_{z}^{(0)}}{\bar{B}_{R}^{(0)}}
\frac{\Delta R}{\bar{h}_{1}}\frac{1}{\Delta\omega' \Delta \Omega}\right)
^{1/2}, 
\end{equation}
and
\begin{equation}
\tau_{3}=\left(\frac{\Delta B_{z}^{(0)}}{\bar{B}_{R}^{(0)}}
\frac{\Delta R}{\bar{h}_{1}}\frac{1}{\Delta\omega' \Delta \omega}\right)
^{1/2}. 
\end{equation}

In order to give guide numbers of these time-scales, let us explore
the limit case in which the amplitude of the warp is sufficiently
high, $h\gg (4\pi\nu_{t}^{2}\hat{\rho})^{-1}B_{R,1}\partial B_{z}/\partial R$,
that the magnetic forces do not affect the vertical motions. In such
case we know from \S \ref{sec:basic} that $\omega=\Omega-\nu_{t}$.
To estimate representative values for $\Delta\Omega$, $\Delta\omega$
and $\Delta \omega'$, we employ the same model than Binney (1992).
Assuming a flat rotation curve and an exponential disc with
scale length $R_{d}$, $\omega (R)$ can be approximated by
\begin{equation}
\omega(R)\simeq 11\left(\frac{R_{d}}{R}\right)^{3}\Omega(R),
\end{equation} 
beyond $4R_{d}$. For
$R_{1}=4R_{d}$ and $R_{2}=6R_{d}$, it follows that $\Delta \omega\simeq
0.4\Delta \Omega$ and $\Delta\omega'\simeq \Delta\Omega$.
Regarding $\Delta \bmath{B}^{0}$, the following upper bounds can be readily
derived using the framework of Battaner \& Jim\'enez-Vicente (1998)
for magnetically-generated warps: $\Delta B_{z}^{(0)}/\Delta B_{R}^{(0)}
\approx \Delta B_{z}^{(0)}/\bar{B}_{R}^{(0)}\approx
\Delta B_{z}^{(0)}/\bar{B}_{\phi}^{(0)}\leq 4$.
The resulting upper values for a galaxy like the Milky Way with
$R_{d}\approx 3.5$ kpc and $v_{c}\approx 220$ km s$^{-1}$ are
\begin{equation}
\tau^{(a)}\,\,\,\,{\rm and}\,\,\,\,\tau^{(b)}
\leq \frac{1}{\tan\theta_{w}}\,\,\,\,{\rm Gyr},
\end{equation}
\begin{equation}
\tau^{(c)}\leq \frac{0.6}{\sqrt{\tan\theta_{w}}}\,\,\,\,{\rm Gyr}.
\end{equation}
Inserting a typical value for $\tan\theta_{w}$ of $\sim 0.2$,
a conservative upper value for the evolutionary time-scale is $\leq 1.3$ Gyr.

In order for $B_{R}\partial B_{z}/\partial R$ 
to preserve its initial wavenumber $m=1$,
$\omega(R)$ must differ only slightly from $\Omega(R)$ at every radii,
i.e.~$v_{z}$ should be very small or zero.
The limit $v_{z}\rightarrow 0$ corresponds to what we call
the ``particular solution''. When $v_{z}$ is small, the induction
equation simplifies to Eqs.~(\ref{eq:inductionR})--(\ref{eq:inductionZ}).

\end{document}